\documentclass[english]{article}

\usepackage[latin9]{inputenc}
\usepackage{geometry}
\usepackage{float}
\usepackage{mathrsfs}
\usepackage{amsmath}
\usepackage{amssymb}
\usepackage{graphicx}
\usepackage{setspace}

\makeatletter

\def\mathscr{\mathcal}
\date{}

\usepackage{babel}

\usepackage{babel}

\usepackage{babel}

\makeatother

\usepackage{babel}
\begin{document}

\title{Explicit solution for vibrating bar\\
 with viscous boundaries and internal damper}

\author{Vojin Jovanovic\\
 Systems, Implementation \& Integration\\
 Smith Bits, A Schlumberger Co.\\
 1310 Rankin Road\\
 Houston, TX 77032\\
 e-mail: fractal97@gmail.com \and Sergiy Koshkin\\
 Computer and Mathematical Sciences\\
 University of Houston-Downtown\\
 One Main Street, \#S705\\
 Houston, TX 77002\\
 e-mail: koshkins@uhd.edu}
\maketitle
\begin{abstract}
We investigate longitudinal vibrations of a bar subjected
to viscous boundary conditions at each end, and an internal damper
at an arbitrary point along the bar's length. The system is described
by four independent parameters and exhibits a variety of behaviors
including rigid motion, super stability/instability and zero damping.
The solution is obtained by applying the Laplace transform to the
equation of motion and computing the Green's function of the transformed
problem. This leads to an unconventional eigenvalue-like problem with
the spectral variable in the boundary conditions. The eigenmodes of the problem 
are necessarily complex-valued and are not orthogonal in the usual inner product. 
Nonetheless, in generic cases we obtain an explicit eigenmode expansion for the response
of the bar to initial conditions and external force. For some special values of parameters 
the system of eigenmodes may become incomplete, or no non-trivial eigenmodes may exist at all.
We thoroughly analyze physical and mathematical reasons for this behavior and explicitly 
identify the corresponding parameter values. In particular, when no eigenmodes exist, 
we obtain closed form solutions. Theoretical analysis is complemented by numerical simulations, 
and analytic solutions are compared to computations using finite elements.

\end{abstract}
\newpage{}

\section{Introduction}

\label{s1}

In this paper we analyze longitudinal vibrations of a bar with dampers
attached at each end as well as at an internal point of the bar. This
type of problem occurs in modeling structures containing shock absorbers
and in control of continuous structures with discrete elements. Mathematically,
the problem reduces to solving the wave equation modified by a Dirac
delta term with viscous boundary conditions. When the boundary conditions
are classical, e.g. the ends are free or clamped, separation of variables
or Laplace transforms reduce the situation to a boundary eigenvalue
problem for a second-order ODE called the Sturm-Liouville problem.
These problems are self-adjoint and admit a complete system of orthogonal
eigenmodes the solution can be expanded into with coefficients determined
from initial values using the orthogonality.

When viscous boundaries are present the Laplace transform leads to
a boundary value problem with the spectral parameter entering boundary
conditions. One still gets a system of eigenmodes, but they are not
orthogonal in the usual inner product, and the eigenvalues are general complex numbers reflecting
the non-self-adjoint nature of the problem. For some critical values
of damping parameters the system of eigenmodes may not be complete,
and even when it is finding the expansion coefficients in terms of
initial data is non-trivial because the eigenmodes are not orthogonal.
Although studied by mathematicians \cite{OrShk,Tamarkin} such problems
and their properties are rather sparsely treated in the engineering
literature, nevertheless see \cite{Hull,PraterSingh,SinghLyonsPrater,UdwSuper} and example 4 in 
\cite[chap 4]{Friedman}.
Hull \cite{Hull} was first to treat a bar with a viscous boundary
at one end and clamped at the other, but he utilized a non-standard
approach to decoupling the equations of motions and provided a response
only for a harmonic driving force. Udwadia \cite{UdwSuper} appears
to be the first to provide a complete closed form solution to this
problem via Laplace transform.

Adding an internal damper at an arbitrary point of the bar, as we
do in this paper, significantly complicates a closed form solution
to the problem. In particular, it is no longer possible to find analytic
formulas for the eigenvalues since their determination depends on
solving algebraic equations of arbitrarily high degree. However, if
the eigenvalues are found numerically the solution can be written
explicitly in a closed form. We obtain the analytical solution by
taking the Laplace transform and finding the Green's function for the
resulting boundary eigenvalue problem. In our analysis we are able
to take advantage of general mathematical results that simplify calculations
considerably. For example, although we do find a cumbersome explicit
formula for the Green's function it is not necessary to find the expansion
of its inverse Laplace transform, which depends on eigenmodes and
eigenvalues only. Moreover, many qualitative traits of the solution
can be gleaned from the characteristic equation for eigenvalues directly
without computing the vibratory response. Internal damper in a problem
with free ends was considered in \cite{Krenk}.

Behavior of the bar is controlled by four dimensionless parameters,
the damping coefficients $h_{1}$ and $h_{2}$ at the left and right
ends, the internal damping coefficient $h_{3}$, and the ratio $a/L$
characterizing the position of the internal damper ($a$ is the distance
to the damper from the left end of the bar, and $L$ is its length).
Since the dimension of the parameter space is four it can not be easily
visualized. As the parameters are varied, the bar exhibits a variety
of behaviors including rigid motions, zero damping, super stability
and instability. Although a four-dimensional diagram can not be drawn
we give analytic conditions for all these types of behavior. Much
of the unusual behavior is due to the fact that we do not restrict
$h_{i}$'s to positive values they would take if the dampers are realized
as dashpots. For the negative values we are dealing with so-called
active dashpots, or rather 'pushpots', that add energy to the bar
instead of damping it. Such discrete elements are sometimes used in
control problems for continuous structures \cite{UdwSuper}.

Perhaps the most striking observation is the extreme sensitivity of
the eigenvalue distribution to the nature of the number $a/L$. When
this number is rational the eigenvalues are generically distributed
along $p$ vertical lines in the complex plane, where $p$ is the
denominator of $a/L$ in lowest terms. This significantly complicates
expansion into eigenmodes since increasingly larger numbers of them
have to be kept. When $a/L$ is irrational this distribution appears random. 
Vibratory response on the other hand, is qualitatively
insensitive to the placement of the internal damper, and in practice 
one may want to use ratios with small denominators like $1/2,1/3,2/3$,
etc. A flip side of these observations is that while FEM produces
good approximation for the vibratory response at least for short times
it performs poorly in approximating the eigenvalues. In fact, it produces
spurious eigenvalues with large real parts that do not converge to
actual eigenvalues as the number of elements is increased. When the
real parts of the spurious eigenvalues are positive FEM will lead
to large errors in the vibratory response at large times.

We organize our presentation as follows. The next section gives
the precise problem statement and describes our approach towards solving
it and the main results obtained. In sections \ref{s3},\ref{s4}
we respectively derive analytic formulas for the eigenmodes, and reduce
computing the eigenvalues to solving an algebraic equation for $a/L$
rational. Section \ref{s5} discusses in more detail the case, when
the damper is placed exactly in the middle of the bar, i.e. $a/L=1/2$.
We compute the eigenvalues explicitly and also give explicit conditions
for the undamped behavior of the bar. The Laplace transform of the
Green's function of our problem is computed in Section \ref{s6} and
we discuss its expansion into partial fractions. Under generic conditions
such expansion exists and is easily Laplace inverted providing a convenient
way for solving our initial-boundary problem. We also give formulas
for special combinations of parameters when the Green's function can
be inverted analytically. Section \ref{s7} presents theoretical analysis
of eigenmode completeness for our problem, and discusses the physical
meaning of critical cases when this completeness is lost. In section
\ref{s8} we use the eigenmode expansion of the Green's function to
write the vibratory response of the bar to initial data and external
force. Section \ref{s8} compares analytic and FEM solutions in several
parameter regimes and discusses spurious eigenvalues produced by FEM.
Finally, we draw our conclusions in section \ref{s9}. Appendix gives
derivations of some formulas used in the main text.

\section{Main results}

\label{s2}

\begin{figure}[h]

\begin{centering}
\includegraphics{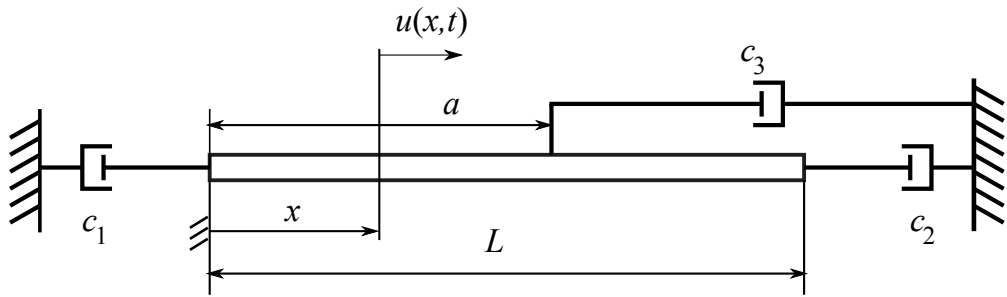} 
\par\end{centering}

\caption{\label{cap:system}A bar with viscous ends and internal damper.}
\end{figure}

We begin with the problem statement. Figure \ref{cap:system} depicts
a bar of length $L$ free to move horizontally suspended by two dampers
at each end and by one at the distance $a$ from its left end. Symbols
$\rho,A$ and $E$ represent the density of the bar, the constant
cross-sectional area and its modulus of elasticity respectively, the
wave speed along the bar is denoted $c:=(E/\rho)^{1/2}$. Let $c_{1}$,
$c_{2}$ and $c_{3}$ be the damping coefficients of the left, right
and internal dampers respectively, we set $h_{1}:=\frac{c}{EA}c_{1}$,
$h_{2}:=\frac{c}{EA}c_{2}$ and $h_{3}:=\frac{c}{2EA}c_{3}$ (the
extra $1/2$ simplifies some formulas). These $h_{i}$ along with
$a/L$ are the dimensionless parameters that determine qualitative
behavior of the bar. Since in our case the bar can move rigidly just
as in the problem with free ends, we write the equation of motion
in the absolute frame that remains at rest at all times. At $t=0$
the left end of the bar is assumed to be at the origin, and $u(x,t)$
denotes the displacement of the point with initial coordinate $x$
at time $t$, see Fig.\ref{cap:system}. 

The system is described by a modified wave equation 
\begin{equation}
u_{tt}(x,t)+2h_{3}c\,\delta(x-a)\, u_{t}(x,t)-c^{2}u_{xx}(x,t)=p(x,t),\label{eq:eom}
\end{equation}
 with the boundary conditions 
\begin{equation}
u_{x}(0,t)-\frac{h_{1}}{c}\, u_{t}(0,t)=0\qquad\mbox{and}\qquad u_{x}(L,t)+\frac{h_{2}}{c}\, u_{t}(L,t)=0.\label{eq:eombc}
\end{equation}
 Here $p(x,t)$ is the external force per unit mass, and the subscripts
$x$, $t$ denote partial derivatives with respect to space and time.
Given $u(x,t)$ the solution in the frame that moves along with the
left end of the bar is $u(x,t)-u(0,t)$.

To solve the problem we use Laplace transforms. Setting the external
force and initial data to zero and taking Laplace transform we get
a homogeneous boundary problem for $U(x,s):=\mathscr{L}[u(x,t)]$
\begin{equation}
\frac{s^{2}}{c^{2}}\, U(x,s)+2h_{3}\frac{s}{c}\,\delta(x-a)\, U(x,s)-U_{xx}(x,s)=0,\label{eq:Leom}
\end{equation}
 
\begin{equation}
U_{x}(0,s)-h_{1}\frac{s}{c}\, U(0,s)=0\qquad\mbox{and}\qquad U_{x}(L,s)+h_{2}\frac{s}{c}\, U(L,s)=0.\label{eq:Leombc}
\end{equation}
We will explicitly compute the Green's function $G(x,\xi,s)$ for this problem and use it to solve 
the original initial-boundary problem.

Unfortunately, the inverse Laplace transform of $G(x,\xi,s)$ can
not be computed in closed form except in special cases. To invert
the Laplace transform for $G(x,\xi,s)$ we use the spectral method.
Note that one gets the same boundary problem by separating variables
and looking for solutions of the form $u(x,t)=\varphi_{a}(x,s)e^{st}$,
where $s$ is the spectral parameter. System \eqref{eq:Leom},\eqref{eq:Leombc}
is almost an ordinary boundary eigenvalue problem except for the presence
of $s$ in the boundary conditions. One expects that it is solvable
only for special values $s_{n}$, the eigenvalues, with $\varphi_{a}(x,s_{n})$
being the vibrational eigenmodes of the bar. This is indeed the case,
and for $a/L=q/p$ (with integers $q,p$ in lowest terms) they can
be grouped into $p$ series $s_{n}^{(k)}$, one for each root of a
degree $p$ algebraic equation. These roots are the only quantities to be computed numerically. Once they
are determined, $s_{n}^{(k)}$ and the eigenmodes can be written explicitly.

We then show that under generic conditions the Green's function $\mathscr{L}^{-1}[G(x,\xi,s)]$
can be expanded into a series over the eigenmodes and give an explicit
formula for the expansion coefficients $A_{n}^{(k)}$. With all the pieces in place we derive our
main result, a series solution for the vibratory response of the bar
with initial data $u(x,0)$, $\dot{u}(x,0)$ subjected to the external
force $p(x,t)$: 
\begin{align}
u(x,t) & =\ \frac{1}{h_{1}+h_{2}+2h_{3}}\Bigg[h_{1}u(0,0)+h_{2}u(L,0)+2h_{3}u(a,0)+\frac{1}{c}\int_{0}^{L}\Bigl[\dot{u}(\xi,0)+\int_{0}^{t}p(\xi,\tau)\, d\tau\Bigr]d\xi\,\Bigg]\notag\label{eq:MRuxt}\\
 & \quad+\sum_{k=1}^{p}\sum_{n=-\infty}^{\infty}\ \frac{\varphi_{a}(x,s_{n}^{(k)})\, e^{s_{n}^{(k)}t}}{c^{2}A_{n}^{(k)}}\Bigg[\ \Bigl[c\, h_{1}\, u(0,0)+c\, h_{2}\, u(L,0)\,\varphi_{a}(L,s_{n}^{(k)})+2c\, h_{3}\, u(a,0)\,\varphi_{a}(a,s_{n}^{(k)})\Bigr]\\
 & \hspace{5.2cm}+\int_{0}^{L}\Bigl[s_{n}^{(k)}\, u(\xi,0)+\dot{u}(\xi,0)+\int_{0}^{t}p(\xi,\tau)\, e^{-s_{n}^{(k)}\tau}\, d\tau\Bigr]\varphi_{a}(\xi,s_{n}^{(k)})\, d\xi\ \Bigg].\notag
\end{align}
 The other major result is a complete analysis of critical behavior.
Namely, we prove that the eigenmodes and associated modes are sufficient
to expand the Green's function if and only if $h_{i}\neq\pm1$. When
$h_{2}=1,h_{3}=0$ or $h_{1}=h_{2}=1$ we derive alternative solutions
in closed form.

\section{Eigenmodes}

\label{s3}

In this section we compute the eigenmodes of our problem analytically.
To aid notation and understanding we first compute the eigenmodes
for the simpler system without the internal damper ($h_{3}=0$). Define
$\varphi(x,s),\ \psi(x,s)$ as solutions to $\frac{s^{2}}{c^{2}}\, U(x,s)-U_{xx}(x,s)=0$
satisfying only the left and the right boundary condition from Eq.\eqref{eq:Leombc}
respectively. This defines them up to a constant multiple and we make
them unique by normalizing $\varphi(0,s)=1=\psi(L,s)$. One easily
finds 
\begin{equation}
\varphi(x,s)=\cosh\Bigl(\frac{sx}{c}\Bigr)+h_{1}\sinh\Bigl(\frac{sx}{c}\Bigr)=\frac{1}{2}\left[(1+h_{1})e^{\frac{sx}{c}}+(1-h_{1})e^{-\frac{sx}{c}}\right]\label{eq:fi}
\end{equation}
 
\begin{equation}
\psi(x,s)=\cosh\Bigl(\frac{s(L-x)}{c}\Bigr)+h_{2}\sinh\Bigl(\frac{s(L-x)}{c}\Bigr)=\frac{1}{2}\left[(1+h_{2})e^{\frac{s(L-x)}{c}}+(1-h_{2})e^{-\frac{s(L-x)}{c}}\right].\label{eq:psi}
\end{equation}
 By construction, any solution to $\frac{s^{2}}{c^{2}}\, U-U_{xx}=0$
satisfying the left (right) boundary condition must be a multiple
of $\varphi(x,s)$ ($\,\psi(x,s)\,$). Since the eigenmodes are supposed
to satisfy both we conclude that they only exist for values of $s$
that make $\varphi(x,s)$ and $\psi(x,s)$ proportional. For such
values either one of them can be taken as the eigenmode. Proportionality
happens if and only if the Wronskian $W[\varphi,\psi]:=\left|\begin{array}{cc}
\varphi & \psi\\
\varphi'_{x} & \psi'_{x}
\end{array}\right|$ is zero giving us an equation for eigenvalues $W[\varphi,\psi]=0$.
For future reference we denote 
\begin{align}
\Delta(s):=-\frac{c}{s}W[\varphi,\psi] & =(1+h_{1}h_{2})\sinh\Bigl(\frac{sL}{c}\Bigr)+(h_{1}+h_{2})\cosh\Bigl(\frac{sL}{c}\Bigr)\notag\label{eq:del}\\
 & =\frac{1}{2}\left[(1+h_{1})(1+h_{2})e^{\frac{sL}{c}}-(1-h_{1})(1-h_{2})e^{-\frac{sL}{c}}\right],
\end{align}
and remark that the eigenvalues other than $s=0$ satisfy $\Delta(s)=0$ or 
$$
e^{\frac{2sL}{c}}=\frac{(1-h_{1})(1-h_{2})}{(1+h_{1})(1+h_{2})}\,.
$$
The latter can be solved explicitly: 
\begin{equation}
s_{n}=\frac{c}{2L}\left[\ln\left|\frac{(1-h_{1})(1-h_{2})}{(1+h_{1})(1+h_{2})}\right|+i\Bigl(\text{Arg}\Bigl(\frac{(1-h_{1})(1-h_{2})}{(1+h_{1})(1+h_{2})}\Bigr)+2\pi n\Bigr)\right],\quad n\in\mathbb{Z}.\label{eq:sn}
\end{equation}
 Substituting $s_{n}$ into Eq.\eqref{eq:fi} gives the eigenmodes
$\varphi(x,s_{n})$, also explicitly.

We now apply the same scheme to Eq.\eqref{eq:Leom} when $h_{3}\neq0$.
As above, denote by $\varphi_{a}(x,s)$ ($\psi_{a}(x,s)$) solutions
to Eq.\eqref{eq:Leom} satisfying the left (right) boundary condition
and normalized to be $1$ at the corresponding boundary. Consider
$\varphi_{a}$ first. Since the damper at $x=a$ does not affect the
equation on $[0,a)$ we have $\varphi_{a}(x,s)=\varphi(x,s)$ on this
interval.

For $x>a$ our $\varphi_{a}$ again satisfies $\frac{s^{2}}{c^{2}}\, U-U_{xx}=0$
but there must be a jump in its first derivative at $a$ to produce
the $2h_{3}\frac{s}{c}\,\delta(x-a)\, U$ term in Eq.\eqref{eq:Leom}.
Along with continuity at $a$ we have $\varphi_{a}(a,s)-\varphi(a,s)=0$
and $\varphi'_{a}(a,s)-\varphi'(a,s)=2h_{3}\frac{s}{c}\varphi(a,s)$.
Since the difference $\varphi_{a}-\varphi$ also satisfies $\frac{s^{2}}{c^{2}}\, U-U_{xx}=0$
we see by inspection that $\varphi_{a}(x,s)=\varphi(x,s)+2h_{3}\,\varphi(a,s)\,\sinh\left(\frac{s(x-a)}{c}\right)$
for $x>a$.

One can compute $\psi_{a}$ analogously or notice that by symmetry
it can be obtained from $\varphi_{a}$ by changing $x$ to $L-x$,
$a$ to $L-a$, and $h_{1}$ to $h_{2}$. With the help of the unit
step Heaviside function $H(x):=\begin{cases}
1,\quad x\geq0\\
0,\quad x<0
\end{cases}$ the $x<a$ and $x>a$ cases can be unified into 
\begin{equation}
\varphi_{a}(x,s)=\varphi(x,s)+2h_{3}\, H(x-a)\,\varphi(a,s)\,\sinh\Bigl(\frac{s(x-a)}{c}\Bigr)\label{eq:fia}
\end{equation}
 
\begin{equation}
\psi_{a}(x,s)=\psi(x,s)+2h_{3}\, H(a-x)\,\psi(a,s)\,\sinh\Bigl(\frac{s(a-x)}{c}\Bigr)\label{eq:psia}
\end{equation}
 To further aid in notation we define 
\begin{equation}
\Delta_{a}(s)=-\frac{c}{s}W[\varphi_{a},\psi_{a}]=\Delta(s)+2h_{3}\varphi(a,s)\psi(a,s)\label{eq:trianglea}
\end{equation}
 and compute explicitly 
\begin{eqnarray}
\Delta_{a}(s) & = & \left(1+h_{1}h_{2}+h_{3}(h_{1}+h_{2})\right)\sinh\Bigl(\frac{sL}{c}\Bigr)+\left(h_{1}+h_{2}+h_{3}(1+h_{1}h_{2})\right)\cosh\Bigl(\frac{sL}{c}\Bigr)\notag\label{eq:dela}\\
 &  & +h_{3}(h_{2}-h_{1})\sinh\Bigl(\frac{s(L-a)}{c}\Bigr)+h_{3}(1-h_{1}h_{2})\cosh\Bigl(\frac{s(L-a)}{c}\Bigr).
\end{eqnarray}
 We will mostly use the exponential form of this expression 
\begin{eqnarray}
\Delta_{a}(s) & = & \frac{1}{2}\left[(1+h_{1})(1+h_{2})(1+h_{3})\, e^{\frac{sL}{c}}-(1-h_{1})(1-h_{2})(1-h_{3})\, e^{-\frac{sL}{c}}\right.\notag\label{eq:delae}\\
 &  & \left.+(1-h_{1})(1+h_{2})\, h_{3}\, e^{\frac{s(L-2a)}{c}}+(1+h_{1})(1-h_{2})\, h_{3}\, e^{-\frac{s(L-2a)}{c}}\right].
\end{eqnarray}
 Eigenvalues are the solutions to $s\Delta_{a}(s)=0$, but $\Delta_{a}(s)$
contains four distinct exponents and $\Delta_{a}(s)=0$ in general
can not be solved explicitly. Still, if the zeros are found numerically
one can get explicit expressions for the eigenmodes from Eq.\eqref{eq:fia}.
Note that $s=0$ corresponds to a rigid displacement of the bar since
$\varphi_{a}(x,0)=1$. We address computation of other eigenvalues
in the next section.

\section{Eigenvalues}

\label{s4}

We now need to compute the eigenvalues, i.e. solve the characteristic
equation $s\Delta_{a}(s)=0$. As we saw in the previous section, when
there is no internal damper the answer is given explicitly by Eq.\eqref{eq:sn}.
This is possible because only two different exponents appear in $\Delta(s)$.
However, in general $\Delta_{a}(s)$ contains four different exponents
and there is no analytic formula for its zeros. Nonetheless, it is
possible to reduce this transcendental equation to solving an algebraic
one. Let us multiply the exponential form of $\Delta_{a}$ by $e^{\frac{sL}{c}}$
and set: 
\[
A_{1}=(1+h_{1})(1+h_{2})(1+h_{3}),\quad A_{2}=h_{3}(1-h_{1})(1+h_{2}),\quad A_{3}=h_{3}(1+h_{1})(1-h_{2}),\quad A_{4}=-(1-h_{1})(1-h_{2})(1-h_{3}).
\]
 In terms of a new variable $\xi:=2sL/c$ the equation becomes 
\begin{equation}
A_{1}\, e^{\xi}+A_{2}\, e^{\left(1-\frac{a}{L}\right)\xi}+A_{3}\, e^{\frac{a}{L}\xi}+A_{4}=0.\label{eq:AiCharEq}
\end{equation}
The left hand side is a so-called exponential sum with real exponents,
its zeros can be distributed in complicated ways. However, these complications
only arise when the exponents in Eq.\eqref{eq:AiCharEq} are incommensurable,
i.e. $a/L$ is irrational \cite[chap VI]{Levin}. But irrational numbers can
be approximated by rational numbers with arbitrary precision, so one
can assume, for all practical purposes, that $a/L$ is rational. We
do so from now on.

Let $a/L=q/p$ in lowest terms with positive integers $p,q$. A further
substitution $z=e^{\frac{\xi}{p}}=e^{\frac{2sL}{pc}}$ reduces Eq.\eqref{eq:AiCharEq}
to an algebraic equation 
\begin{equation}
A_{1}z^{p}+A_{2}z^{p-q}+A_{3}z^{q}+A_{4}=0.\label{eq:AlgCharEq}
\end{equation}
If $A_{1}\neq0$ it has $p$ roots $z_{1},\, z_{2},...,\, z_{p}$
counting multiplicity, and we need not deal with infinitely many zeros
of Eq.\eqref{eq:AiCharEq} directly. If moreover $A_{4}\neq0$ none
of these roots is $0$. When $A_{1}$ or $A_{4}$ do vanish we get
critical cases that require special treatment.

\begin{figure}[h]
\begin{centering}
\includegraphics[scale=0.6]{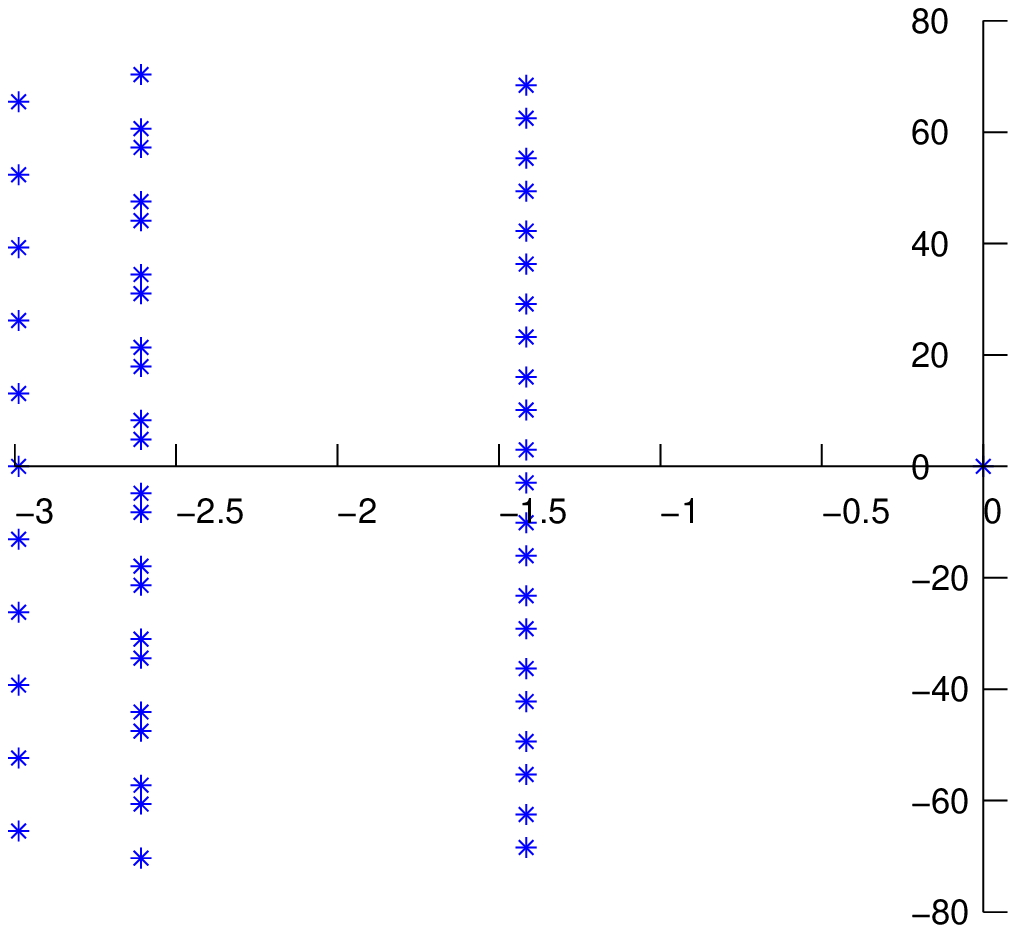}\includegraphics[scale=0.6]{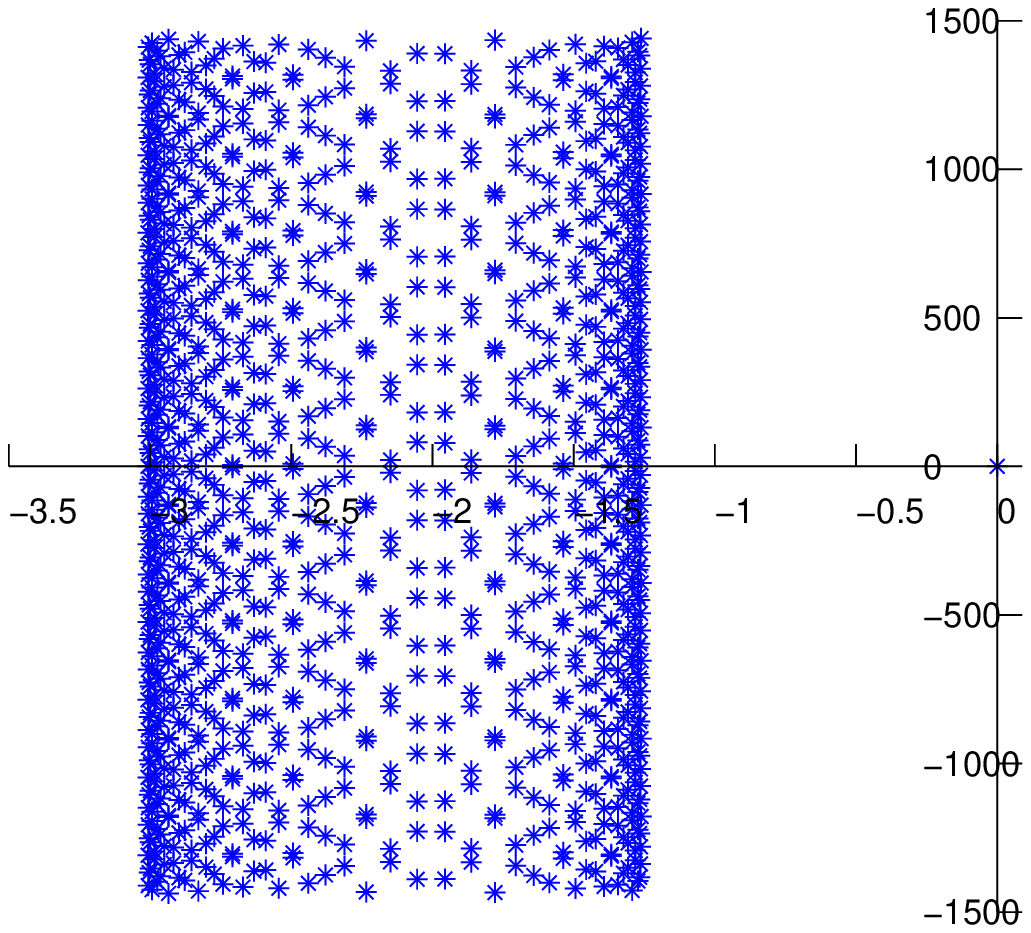} 
\par\end{centering}
\caption{Distribution of eigenvalues for $a/L=2/5$ (left), and $a/L=41/100$
(right) with $h_{1}=0.3$, $h_{2}=0.9$, $h_{3}=0.7$, $c=0.3$ and $L=1.8$.
Horizontal lines are the real axes, vertical lines are parallel to
the imaginary axes.}
\label{cap:eigenvalues} 
\end{figure}

In general, roots $z_{k}$ can not be found analytically as functions
of $h_{i}$. However, there is a standard way of setting up a matrix
with the characteristic equation Eq.\eqref{eq:AlgCharEq} so that
they can be found numerically as its eigenvalues using MATLAB or Maple.
Once $z_{k}$ are found, the eigenvalues can be expressed as (cf. Eq.\eqref{eq:sn})
\begin{equation}
s_{n}^{(k)}=\frac{p\, c}{2L}\left[\,\ln\left|z_{k}\right|+i\left(\text{Arg}\left(z_{k}\right)+2\pi n\right)\,\right],\quad n\in\mathbb{Z},\ k=1,\dots,p\,.\label{eq:snk}
\end{equation}
Fig.\ref{cap:eigenvalues} shows a typical distribution of eigenvalues, 
MATLAB function eig() was used to compute the roots $z_{k}$.

If all $z_{k}$ are distinct then $s_{n}^{(k)}$ fall into $p$ infinite
sequences with $\text{Re}(s)=\frac{pc}{2L}\ln\left|z_{k}\right|$
equispaced along vertical lines in the complex plane. The lines are
not necessarily distinct even if the roots are, since the latter may
have equal absolute values. In fact, if Eq.\eqref{eq:AlgCharEq} has
a complex root $z_{k}$ then its conjugate $\overline{z}_{k}$ is
also a root, and the corresponding line contains two different sequences
$s_{n}^{(k)}$ with arguments of opposite signs. The eigenvalues are
spaced twice as densely along such lines, see Fig.\ref{cap:eigenvalues}.
If more roots have the same absolute value the density increases further,
but the eigenvalues remain distinct as long as the roots are.

However, if Eq.\eqref{eq:AlgCharEq} does have multiple roots then
according to Eq.\eqref{eq:snk} we get an infinite sequence of eigenvalues
with the same multiplicity. Multiple eigenvalues create extra difficulties
in computing the Green's function since the eigenmodes are no longer
sufficient to express the response (the associated modes also enter).
We shall assume henceforth that all the roots of Eq.\eqref{eq:AlgCharEq}
are simple. Then all the eigenvalues are also simple with the possible
exception of $s=0$, which will have multiplicity two if $\Delta_{a}(0)=h_{1}+h_{2}+2\, h_{3}=0$.
It corresponds to a bar moving as a rigid body with constant velocity.

When $p$ is small $z_{k}$ can be found in closed form. The simplest
such case is $p=2$, $q=1$ so that $a=\frac{L}{2}$, i.e. the internal
damper is placed exactly in the middle of the interval $[0,L]$. We
wish to analyze this case in more detail next.

\section{Damper in the middle\label{sec:Damper-in-the}}\label{s5}

When $a/L=1/2$ Eq.\eqref{eq:AlgCharEq} reduces to a quadratic equation:
\begin{equation}
Az^{2}+Bz+C=0\quad\text{with}\label{eq:QCharEq}
\end{equation}
\[
A=(1+h_{1})(1+h_{2})(1+h_{3}),\quad B=2h_{3}(1-h_{1}h_{2}),\quad C=-(1-h_{1})(1-h_{2})(1-h_{3}).
\]
As above, we exclude critical cases so that $A\neq0$ and $C\neq0$.
Then Eq.\eqref{eq:QCharEq} has two real roots, two complex conjugate
roots or one real double root according to whether 
\begin{equation}
D:=\frac{B^{2}-4AC}{4}=(1-h_{1}^{2})(1-h_{2}^{2})+h_{3}^{2}\,(h_{1}-h_{2})^{2}\label{eq:Discrim}
\end{equation}
is positive, negative or zero respectively. Moreover, we have explicitly
\begin{equation}
z_{1,2}=\frac{-h_{3}(1-h_{1}h_{2})\pm\sqrt{D}}{(1+h_{1})(1+h_{2})(1+h_{3})}.\label{eq:Qroots}
\end{equation}
Accordingly, for $D>0$ we have two sequences of eigenvalues spaced
along two distinct vertical lines, see Eq.\eqref{eq:snk}. They merge
into a single line of double eigenvalues for $D=0$, and for $D<0$
we have both sequences half-spaced along a single line with 
\[
\text{Re}(s)=\frac{c}{4L}\,\ln\left|z_{1,2}\right|=\frac{c}{8L}\ln\left|\frac{(1-h_{1})(1-h_{2})(1-h_{3})}{(1+h_{1})(1+h_{2})(1+h_{3})}\right|.
\]
Even if $D\neq0$ when $h_{1}+h_{2}+2\, h_{3}=0$ we have a double
eigenvalue at zero because of the rigid mode, in which case $D=(h_{1}^{2}+h_{2}^{2}-2)^{2}\neq0$.

As an application, we will determine conditions for 'undamped' behavior
of the bar. The idea is this: when $h_{i}$'s are negative they describe
active dashpots that add energy to the bar instead of draining
it \cite{UdwSuper}. It is therefore possible that for some combinations
of values the same amount of energy is being added as is being drained
by the dashpots -- the bar behaves as if it were undamped altogether.
Clearly, no damping means that all the eigenvalues are purely imaginary
or equivalently, that all the roots of Eq.\eqref{eq:QCharEq} lie
on the unit circle since $z=e^{\frac{sL}{c}}$.

If both roots are real then $z=1,1$, $z=-1,-1$ or $z=1,-1$. This
leads to $C=A,\ B=\pm2A$ or $C=-A,\ B=0$ respectively. If one of
them is complex then its conjugate is also a root and $z=e^{\pm i\theta}$
with $\theta\neq0,\pi$. We have 
\[
Az^{2}+Bz+C=A(z-e^{\theta})(z-e^{-\theta})=Az^{2}-2A\cos\theta+A,
\]
 so $C=A$ and $\cos\theta=-B/2A$. The latter condition can be satisfied
by a $\theta\neq0,\pi$ if and only if $|B|<2|A|$. Two of the real
cases above will be subsumed here if we allow $|B|\leq2|A|$.

To make these conditions explicit in terms of $h_{i}$'s let us start
with the case $h_{3}=0$. This automatically means $B=0$, in particular
$|B|\leq2|A|$, so $C=A$ and $C=-A$ give us the two available possibilities:
\[
\textbf{(1) }\begin{cases}
h_{3}=0\\
h_{1}h_{2}=-1
\end{cases}\text{and}\qquad\textbf{(2) }\begin{cases}
h_{3}=0\\
h_{1}+h_{2}=0.
\end{cases}
\]
 In each case $h_{1}$ or $h_{2}$ can be chosen arbitrarily and the
other parameter is uniquely defined. When $h_{3}\neq0$ we have a
similar situation for $C=-A$: 
\[
\textbf{(3) }\begin{cases}
h_{3}=-\frac{h_{1}+h_{2}}{2}\\
h_{1}h_{2}=1.
\end{cases}
\]
 Note that in the second and third case we have $h_{1}+h_{2}+2h_{3}=0$,
which means that there is a double pole at $s=0$. Therefore, we have
not just zero damping but rigid motion: the bar will be moving with
constant velocity in addition to oscillations.

When $h_{3}\neq0$ the case of $C=A$ becomes unexpectedly complicated.
After cancellations $C=A$ reduces to $h_{1}h_{2}+h_{1}h_{3}+h_{2}h_{3}+1=0$
and we also get 
\[
\cos\theta=-\frac{1-h_{1}^{2}h_{2}^{2}}{(1-h_{1}^{2})(1-h_{2}^{2})}
\]
 We can solve for $h_{3}$ since $h_{1}+h_{2}=0$ leads to one of
the above cases, but writing out $|B|\leq2|A|$ is not very helpful
for determining $h_{1},h_{2}$. We will assume instead that one picks
$h_{1}$ and $\cos\theta$ and then solves for $h_{2},h_{3}$: 
\[
\textbf{(4) }\begin{cases}
h_{3}=-\frac{1+h_{1}h_{2}}{h_{1}+h_{2}}\\
h_{2}=\pm\sqrt{\frac{(1+\cos\theta)-h_{1}^{2}\cos\theta}{\cos\theta+(1-\cos\theta)h_{1}^{2}}}.
\end{cases}
\]
One can see that the range of parameters in case \textbf{(4)} is two-dimensional
in contrast to the first three cases. Still, the choice of $h_{1}$
and $\cos\theta$ is not entirely arbitrary. Aside from the obvious
restriction $-1\leq\cos\theta\leq1$ one has to pick $h_{1}$ so that
the expression under the square root is non-negative. For example,
if $\cos\theta=1$ then $h_{2}=\pm\sqrt{2-h_{1}^{2}}$ and we can
only pick $h_{1}$ that satisfies $|h_{1}|\leq\sqrt{2}$. In particular,
$|h_{1}|\leq1$ always works for $\cos\theta>0$, and $|h_{1}|\geq1$
for $\cos\theta<0$.

For $C=A$ and $B=\pm2A$, which corresponds to $\cos\theta=\pm1$,
we have multiplicity in eigenvalues. Indeed, Eq.\eqref{eq:snk} shows
that all eigenvalues on the imaginary axis are now double zeros of
$\Delta_{a}(s)$. Therefore, we have oscillations with linearly increasing
amplitudes. Moreover, when $\cos\theta=1$ we have that $0$ is even
a triple pole of $s\Delta_{a}(s)$ and hence a triple eigenvalue.
Physically, this means that the bar does not just move rigidly, it
is accelerating.

\section{Green's function}

\label{s6}

By definition, the Green's function $G(x,\xi,s)$ for system \eqref{eq:Leom}-\eqref{eq:Leombc}
satisfies 
\begin{equation}
\frac{s^{2}}{c^{2}}\, G+2h_{3}\frac{s}{c}\,\delta(x-a)\, G-G_{xx}(x,s)=\delta(x-\xi),\label{eq:Geom}
\end{equation}
 
\begin{equation}
G_{x}(0,\xi,s)-h_{1}\frac{s}{c}\, G(0,\xi,s)=0\qquad\mbox{and}\qquad G_{x}(L,\xi,s)+h_{2}\frac{s}{c}\, G(L,\xi,s)=0.\label{eq:Geombc}
\end{equation}

It can be computed along the same lines as $\varphi_{a},\psi_{a}$
in section \ref{s3}, and has different analytic expressions depending
on relative positions of $x$, $a$ and $\xi$. Consider the case
$a<\xi$ first. As a function of $x$, $G$ satisfies Eq.\eqref{eq:Leom}
for $x<\xi$ and $x>\xi$. Therefore, it is equal to $A\varphi_{a}(x,s)$
on $[0,\xi)$ and $B\psi(x,s)$ on $(\xi,L]$. At $x=\xi$ it is continuous,
but has a jump in the first derivative to produce $\delta(x-\xi)$
in Eq.\eqref{eq:Geom}. Namely, $G_{x}(\xi^+,\xi,s)-G_{x}(\xi^-,\xi,s)=-1$
because $G_{xx}$ enters Eq.\eqref{eq:Geom} with minus. Therefore,
$A,B$ can be found from the system 
\[
\begin{cases}
A\varphi_{a}(\xi,s)-B\psi(\xi,s) & =0\\
A\varphi'_{a}(\xi,s)-B\psi'(\xi,s) & =1
\end{cases}
\]
 Solving for them in the matrix form we get 
\[
\begin{pmatrix}A\\
B
\end{pmatrix}=\frac{1}{-W[\varphi_{a},\psi]}\begin{pmatrix}-\psi' & \psi\\
-\varphi'_{a} & \varphi_{a}
\end{pmatrix}\begin{pmatrix}0\\
1
\end{pmatrix}
\]
 But for $\xi>a$ we have from Eq.\eqref{eq:psia} that $\psi_{a}(\xi,s)=\psi(\xi,s)$
so that $W[\varphi_{a},\psi_{a}]=W[\varphi_{a},\psi]=-\frac{c}{s}\Delta_{a}(s)$
from Eq.\eqref{eq:trianglea}. Therefore, 
\[
A=c\,\frac{\psi(\xi,s)}{s\Delta_{a}(s)}\qquad\text{and}\qquad B=c\,\frac{\varphi_{a}(x,s)}{s\Delta_{a}(s)},
\]
 so that 
\begin{equation}
G_{a<\xi}(x,\xi,s)=\frac{c}{s\Delta_{a}(s)}\begin{cases}
\varphi_{a}(x,s)\psi(\xi,s),\ x<\xi\\
\psi(x,s)\varphi_{a}(\xi,s),\ x>\xi
\end{cases}\label{eq:Galxi}
\end{equation}
Analogously, for $a>\xi$ in the latter case we have 
\begin{equation}
G_{a>\xi}(x,\xi,s)=\frac{c}{s\Delta_{a}(s)}\begin{cases}
\varphi(x,s)\psi_{a}(\xi,s),\ x<\xi\\
\psi_{a}(x,s)\varphi(\xi,s),\ x>\xi
\end{cases}\label{eq:Gagxi}
\end{equation}
It will be convenient for us to rewrite $G$ in a form that is both
more explicit, and makes its symmetry $G(x,\xi,s)=G(\xi,x,s)$ manifest.
To this end, we introduce 
$$
g_{\varphi\psi}(x,\xi,s):=\begin{cases}
\varphi(x,s)\psi(\xi,s),\ x<\xi\\
\varphi(\xi,s)\psi(x,s),\ x>\xi
\end{cases}\!\!\!\!\!\!,
$$ 
and compute 
\begin{multline}\label{eq:fipsi}
g_{\varphi\psi}(x,\xi,s)=\frac{1}{4}\left[(1+h_{1})(1+h_{2})e^{\frac{s(L-|x-\xi|)}{c}}+(1-h_{1})(1+h_{2})e^{\frac{s(L-(x+\xi))}{c}}\right.\\
\left.+(1+h_{1})(1-h_{2})e^{-\frac{s(L-(x+\xi))}{c}}+(1-h_{1})(1-h_{2})e^{-\frac{s(L-|x-\xi|)}{c}}\right].
\end{multline}
Analogously, let
$$
g_{s\psi}(x,\xi,s):=\begin{cases}
\sinh\frac{s(x-a)}{c}\,\psi(\xi,s),\ x<\xi\\
\sinh\frac{s(\xi-a)}{c}\,\psi(x,s),\ x>\xi
\end{cases}\quad\text{and}\quad 
g_{s\varphi}(x,\xi,s):=\begin{cases}
\sinh\frac{s(a-\xi)}{c}\,\varphi(x,s),\ x<\xi\\
\sinh\frac{s(a-x)}{c}\,\varphi(\xi,s),\ x>\xi\,.
\end{cases}
$$ 
Then 
\begin{multline}\label{eq:spsi}
g_{s\psi}(x,\xi,s)=\frac{1}{4}\left[(1+h_{2})e^{\frac{s(L-a-|x-\xi|)}{c}}-(1+h_{2})e^{\frac{s(L+a-(x+\xi))}{c}}\right.\\
\left.+(1-h_{2})e^{-\frac{s(L+a-(x+\xi))}{c}}-(1-h_{2})e^{-\frac{s(L-a-|x-\xi|)}{c}}\right]
\end{multline}
\begin{multline}\label{eq:sfi}
g_{s\varphi}(x,\xi,s)=\frac{1}{4}\left[(1+h_{1})e^{\frac{s(a-|x-\xi|)}{c}}-(1+h_{1})e^{\frac{s((x+\xi)-a)}{c}}\right.
\left.+(1-h_{1})e^{-\frac{s((x+\xi)-a)}{c}}-(1-h_{1})e^{-\frac{s(a-|x-\xi|)}{c}}\right].
\end{multline}
Since the $g_{\varphi\psi}$ part is common to all arrangements of $x$,
$a$ and $\xi$ we get 
\begin{align}
G(x,\xi,s)=\frac{c}{s\Delta_{a}(s)}\Bigl[g_{\varphi\psi}(x,\xi,s)\Bigr. & +2h_{3}\, H(x-a)H(\xi-a)\,\varphi(a,s)\, 
g_{s\psi}(x,\xi,s)\notag\label{eq:eGreen}\\
 & \ \Bigl.+\,2h_{3}\, H(a-x)H(a-\xi)\,\psi(a,s)\, g_{s\varphi}(x,\xi,s)\Bigr].
\end{align}
Note that the last two terms are non-zero only when $x$ and $\xi$
are on the same side of $a$. Therefore, whenever $a$ separates $x$
and $\xi$ the Green's function reduces to the first term.

As mentioned earlier, to solve the original problem we need to invert
the inverse Laplace transform $\Gamma(x,\xi,t):=\mathscr{L}^{-1}[G(x,\xi,s)]$,
but $G$ is too complicated to allow inversion in a closed form. We
are forced to expand it into a series over functions with simpler
dependence on $s$ and invert it termwise. In the spectral method,
which we follow here, these functions are the partial fractions $\frac{1}{(s-p)^{m}}$
and $\mathscr{L}^{-1}\left[\frac{1}{(s-p)^{m}}\right]=\frac{t^{m-1}}{(m-1)!}\, e^{pt}$,
where $p$ are the poles of $G$. Since the denominator of $G$ is
$s\Delta_{a}(s)$ its poles are exactly the eigenvalues $s_{n}^{(k)}$
from Eq.\eqref{eq:snk} and $0$ (none of them cancel with the numerator).

Let us assume for the moment that expansion into partial fractions
is possible and moreover, that all the poles except possibly $s=0$
are simple. A general theorem on Green's functions \cite[chap 1, sect 3]{Naimark} implies
that 
\begin{equation}
G(x,\xi,s)=G_{0}(x,\xi,s)+\sum_{s_{n}^{(k)}\neq0}\frac{\varphi_{a}(x,s_{n}^{(k)})\,\varphi_{a}(\xi,s_{n}^{(k)})}{A_{n}^{(k)}(s-s_{n}^{(k)})},\label{eq:G_expansion}
\end{equation}
 where $G_{0}$ is the principal part at $s=0$, $\varphi_{a}(x,s_{n}^{(k)})$
are the eigenmodes corresponding to $s_{n}^{(k)}$, and $A_{n}^{(k)}$
are numerical coefficients. As shown in the Appendix, in our case
the latter can be computed explicitly: 
\begin{equation}
A_{n}^{(k)}=\frac{2s_{n}^{(k)}}{c^{2}}\int_{0}^{L}\varphi_{a}(\xi,s_{n}^{(k)})^{2}\, d\xi+\frac{h_{1}}{c}\varphi_{a}(0,s_{n}^{(k)})^{2}+\frac{h_{2}}{c}\varphi_{a}(L,s_{n}^{(k)})^{2}+\frac{2h_{3}}{c}\varphi_{a}(a,s_{n}^{(k)})^{2}.\label{Ank}
\end{equation}
If $s=0$ is also a simple pole, i.e. $\Delta_{a}(0)\neq0$, then $G_{0}(x,\xi,s)=cF(x,\xi,0)/s\Delta_{a}(0)$,
where $F$ is the bracketed expression in Eq.\eqref{eq:eGreen}. By
inspection, from Eqs.\eqref{eq:fi},\eqref{eq:psi},\eqref{eq:fia},\eqref{eq:psia}
we see that $\varphi=\psi=\varphi_{a}=\psi_{a}=1$ for $s=0$. Recall
also that $\Delta_{a}(0)=h_{1}+h_{2}+2h_{3}$. Therefore, $F(x,\xi,0)=1$
and 
\begin{equation}
G_{0}(x,\xi,s)=\frac{c}{\left(h_{1}+h_{2}+2h_{3}\right)s}\label{eq:G0}
\end{equation}
The Laplace transform is now easily inverted and 
\begin{equation}
\Gamma(x,\xi,t):=\mathscr{L}^{-1}[G(x,\xi,s)]=\ \frac{c}{h_{1}+h_{2}+2h_{3}}\ +\ \sum_{k=1}^{p}\sum_{n=-\infty}^{\infty}\ \frac{1}{A_{n}^{(k)}}\,\varphi_{a}(x,s_{n}^{(k)})\,\varphi_{a}(\xi,s_{n}^{(k)})\, e^{s_{n}^{(k)}t}\label{eq:Gamt}
\end{equation}

If $s=0$ is a double eigenvalue, i.e. a simple zero of $\Delta_{a}$,
the answer is more cumbersome. For example, when $h_{3}=-\frac{h_{1}+h_{2}}{2}$
and $h_{2}=\frac{1}{h_{1}}$ all the eigenvalues are on the imaginary
axis and there is a double pole at zero. The principal part of the Green's
function at $s=0$ is provided with $G_{0}(x,\xi,s)=\frac{c_{1}}{s}+\frac{c_{2}}{s^{2}}$
where $c_{1}$and $c_{2}$ are computed in the usual way \cite[chap 5, sect 5-11]{LaPage}.
If, for convenience, we set 
\begin{equation}
H_{a}(x,\xi):=H(x-a)H(\xi-a)+H(a-x)H(a-\xi)=\begin{cases}
1,\ \text{\ensuremath{a} is on the same side of \ensuremath{x} and \ensuremath{\xi}}\\
0,\ \text{\ensuremath{a} separates \ensuremath{x} and \ensuremath{\xi}}
\end{cases},
\end{equation}
 then $c_{1}$and $c_{2}$ are provided as 
\begin{align}
c_{1} & =\frac{h_{1}c}{2\left(L(h_{1}^{2}-1)-a(h_{1}^{4}-1)\right)}\Bigg[(h_{1}^{2}+1)\left(|x-\xi|-(x+\xi)+2a\right)H_{a}(x,\xi)\\
 & \hspace{5.5cm}-(h_{1}^{2}+1)|x-\xi|+(h_{1}^{2}-1)(x+\xi)+2L\Bigg]\,,\notag\\
c_{2} & =\frac{h_{1}^{2}c^{2}}{L(h_{1}^{2}-1)-a(h_{1}^{4}-1)}\ .
\end{align}
 The corresponding terms in $\Gamma(x,\xi,t)$ are $c_{1}+c_{2}t$.
Physically, the bar is moving as a rigid body with constant velocity
in addition to vibrating. When $s=0$ has higher multiplicity the bar
will be accelerating. In the case when infinitely many eigenvalues
are multiple, the simple template Eq.\eqref{eq:G_expansion} for the
partial fraction expansion no longer applies. We investigate when
such expansion is possible at all next.

\section{Critical cases}

\label{s7}

Through the previous section we assumed that the Green's function can
be expanded into partial fractions over its poles. This is not always
the case as observed already in \cite{UdwSuper}. In general, if one
takes a function like $e^{-s}$ with no poles at all, it will have no 
expansion in terms of partial fractions.
The usual way of proving convergence is to apply the Cauchy residue
theorem to circles of increasing radii. But for this argument to work,
the contour integrals over the circles must tend to zero as the radii
tend to infinity. This is violated for functions like $e^{-s}$. To
ensure the desired convergence it suffices, for example, that $|G(x,\xi,s)|\leq\text{const}/|s|$
outside disks of any fixed size surrounding the poles of $G$ (the
constant will depend on the size chosen). When this inequality does
not hold we call the case critical.

According to Eq.\eqref{eq:eGreen}, $G(x,\xi,s)=\frac{cF(s)}{s\Delta_{a}(s)}$
with $F(s)$ being the expression in brackets. Since $1/s$ factor
is already present it would suffice that $F(s)/\Delta_{a}(s)$ be
bounded away from the poles. A quick look at explicit expressions
for $F$ and $\Delta_{a}$ shows both to be exponential sums of the
type $a_{1}e^{\alpha_{1}s}+\dots+a_{m}e^{\alpha_{m}s}$ with real
exponents $\alpha_{i}$. Quite a bit is known about such sums, e.g.
their zeros are located within a vertical strip $|\text{Re}(s)|\leq\omega$,
and within this strip the sum is bounded. Moreover, on the complement
to all disks of any fixed size surrounding the zeros exponential sums
are uniformly separated from $0$ \cite{Levin}. Consequently, we
only need to worry about $F(s)/\Delta_{a}(s)$ being bounded when
$\text{Re}(s)\to\pm\infty$.

Clearly, for $\text{Re}(s)\to\infty$ ($-\infty$) the term with the
largest (smallest) $\alpha_{i}$ dominates an exponential sum. We
conclude that $F/\Delta_{a}$ is uniformly bounded away from the poles
if and only if the largest (smallest) exponent in the denominator
is greater (less) than the largest (smallest) exponent in the numerator.
To put it differently: for boundedness all exponents in the numerator
must lie between the largest and the smallest ones in the denominator.
Let us take count of these exponents in Eq.\eqref{eq:eGreen} for
$\xi>a$, the other case is analogous. We find that in the numerator
the following terms occur 
\begin{gather}
e^{\frac{s}{c}(L-|x-\xi|)},\quad e^{\frac{s}{c}(L-(x+\xi)},\quad e^{-\frac{s}{c}(L-|x-\xi|)},\quad e^{-\frac{s}{c}(L-(x+\xi))}\label{eq:NumExps}\\
e^{\frac{s}{c}(L-2a-|x-\xi|)},\quad e^{\frac{s}{c}(L+2a-(x+\xi))},e^{-\frac{s}{c}(L-2a-|x-\xi|)},\quad e^{-\frac{s}{c}(L+2a-(x+\xi))},
\end{gather}
 and the terms in the second row are multiplied by $0$ when $x<a$.
On the other hand, $\Delta_{a}$ contains 
\begin{gather}
e^{\frac{s}{c}L},\quad e^{\frac{s}{c}(L-2a)},\quad e^{-\frac{s}{c}L},\quad e^{-\frac{s}{c}(L-2a)}.\label{eq:DenExps}
\end{gather}
 We see that the exponents of Eq.\eqref{eq:NumExps} do lie between
$-L$ and $L$, so the condition for boundedness is satisfied assuming
that coefficients in front of $e^{\frac{s}{c}L}$ and $e^{-\frac{s}{c}L}$
in $\Delta_{a}(s)$ are non-zero. By Eq.\eqref{eq:delae} this means
that $(1+h_{1})(1+h_{2})(1+h_{3})\neq0$ and $(1-h_{1})(1-h_{2})(1-h_{3})\neq0$,
or equivalently $h_{i}\neq\pm1$ for $i=1,2,3$. For ODE of any order
with linear boundary conditions for non-criticality are derived in
\cite{Tamarkin}.

When say $h_{2}$ is $1$, the only surviving exponents in Eq.\eqref{eq:DenExps}
are the first and the third. This changes the calculation since some
of the exponents in the numerator, like $L-(x+\xi)$, are perfectly
capable of being less than $L-2a$. One may hope that 'bad' terms
in the numerator also vanish, but no, for $h_{2}=1$ the first two
exponents from each row of Eq.\eqref{eq:NumExps} have non-zero coefficients.

When $h_{i}=\pm1$ for at least one $i$ the spectral method no longer
applies and one has to look for other ways to invert the Laplace transform.
Before discussing them let us explain physical reasons for critical
behavior. Take $h_{2}=1$ and note that in the original initial-boundary
problem the right boundary condition becomes $u_{t}(L,t)+c\, u_{x}(L,t)=0$,
while the equation away from the internal damper is $u_{tt}-c^{2}\, u_{xx}=0$.
It is well-known that its solution splits into left and right traveling
waves which satisfy $u_{t}-c\, u_{x}=0$ and $u_{t}+c\, u_{x}=0$
respectively. The left traveling waves play no role near the right
boundary (there is nowhere for them to come from), while the right
traveling waves satisfy the boundary condition automatically. In other
words, the right boundary condition has no effect whatsoever, and
we effectively get a problem on the semi-infinite interval $[0,\infty)$,
but with initial data restricted to $[0,L]$. Physically, the right
boundary becomes transparent allowing the waves to pass through it
without any reflection back. If in addition there is no internal damper
($h_{3}=0$) then standing waves, a.k.a. eigenmodes, can not form,
so of course one can not expand over them. When the internal damper
is present, it reflects some of the waves at $a$ forming standing
waves on $[0,a]$, but not on $[a,L]$. The expansion is still impossible
even though some eigenmodes are present. More physical analysis of
transparent boundaries is given in \cite{UdwSuper}.

When $h_{2}=-1$ the situation can be analyzed similarly. Now the
boundary condition is $u_{t}(L,t)-c\, u_{x}(L,t)=0$ while the equation
prescribes $u_{t}+c\, u_{x}=0$. This is only possible if $u_{t}(L,t)=\, u_{x}(L,t)=0$,
conditions impossible to satisfy in general. If the initial disturbance
is localized away from the right boundary the solution will exist
only up to a time necessary for the right traveling waves to reach
it. As $h_{2}$ approaches $-1$ the reflection coefficient at the
right boundary grows without bound \cite{UdwSuper}, so the physical
reason for the non-existence is that reflection would involve an infinite
energy transfer. One can not say however, that solution blows up in
finite time since it behaves 'normally' before the boundary is reached
and ceases to exist at that instant.

How does one invert the Laplace transform in critical cases? Suppose
first that there is no internal damper ($h_{3}=0$) and the right
boundary is transparent ($h_{2}=1$). Then there is only one exponent
left in the denominator of the Green's function $\Delta_{a}(s)=(1+h_{1})e^{\frac{sL}{c}}$.
This exponent can be combined with the ones in the numerator so that
the Green's function reduces to a linear combination of exponents divided
by $s$ 
\begin{equation}
G(x,\xi,s)=\frac{c}{2s}\left[\, e^{-\frac{s}{c}|x-\xi|}+\frac{1-h_{1}}{1+h_{1}}\, e^{-\frac{s}{c}(x+\xi)}\right].\label{eq:Grh30h21}
\end{equation}
 Since $\mathscr{L}^{-1}[\frac{1}{s}e^{-\alpha s}]=H(t-\alpha)$,
where $H$ is the Heaviside function, $\Gamma(x,\xi,t):=\mathscr{L}^{-1}[G(x,\xi,s)]$
can even be found in closed form: 
\begin{equation}
\Gamma(x,\xi,t)=\frac{c}{2}\left[H(ct-|x-\xi|)+\frac{1-h_{1}}{1+h_{1}}\, H(ct-(x+\xi))\right].\label{eq:Gmh30h21}
\end{equation}
 Physically, we get left and right traveling waves with the former
reflected from the left boundary just once. Thus, it is not surprising
that the solution is a finite superposition of traveling waves, just
as it is for the wave equation on the entire line according to the
d'Alembert's solution.

Summarizing, for $h_{3}=0$ we have a simple dichotomy: either both
boundaries are non-transparent and the solution can be found by the
spectral method, or one or both are and the solution can be found
in a closed form as a finite superposition of traveling waves. When
$h_{3}\neq0$ this dichotomy fails because standing waves may be able
to form on part of the interval between one of the boundaries and
the internal damper. Then one is forced to either combine traveling
and standing waves, or to use an infinite number of traveling waves.
We shall not treat such intermediate cases in this paper.

However, if both boundaries are transparent ($h_{1}=h_{2}=1$) the
standing waves can not form at all and one can find a closed form
solution again. Then from Eq.\eqref{eq:eGreen} 
\begin{align}
G(x,\xi,s) & =\frac{c}{2(1+h_{3})s}\Bigl[e^{-\frac{s}{c}|x-\xi|}+h_{3}H_{a}(x,\xi)\left(e^{-\frac{s}{c}|x-\xi|}-e^{-\frac{s}{c}|x+\xi-2a|}\right)\Bigr];\label{eq:Grh1h21}\\
\Gamma(x,\xi,t) & =\frac{c}{2(1+h_{3})}\Bigl[\, H(ct-|x-\xi|)+h_{3}H_{a}(x,\xi)\,\Bigl(H(ct-|x-\xi|)-H(ct-|x+\xi-2a|)\,\Bigr)\,\Bigr].\label{eq:Gmh1h21}
\end{align}

\section{Vibratory response}

\label{s8}

When computing the Green's function we set all the initial-boundary
data to zero and it is now time to bring it back. The Laplace transform
of the original system \eqref{eq:eom},\eqref{eq:eombc} is 
\begin{equation}
\frac{s^{2}}{c^{2}}\, U(x,s)+2h_{3}\frac{s}{c}\,\delta(x-a)\, U(x,s)-U_{xx}(x,s)=s\,\frac{u(x,0)}{c^{2}}+\frac{\dot{u}(x,0)+p(x,s)}{c^{2}}+2\frac{h_{3}}{c}\,\delta(x-a)\, u(x,0),\label{eq:FLeom}
\end{equation}
 
\begin{equation}
U_{x}(0,s)-h_{1}\frac{s}{c}\, U(0,s)=-\frac{h_{1}}{c}u(0,0)\qquad\mbox{and}\qquad U_{x}(L,s)+h_{2}\frac{s}{c}\, U(L,s)=\frac{h_{2}}{c}u(L,0).\label{eq:FLeombc}
\end{equation}
 If not for the inhomogeneity in the boundary conditions we could
have solved Eqs.\eqref{eq:FLeom},\eqref{eq:FLeombc} by convolution
with the Green's function. Let us denote this convolution $\widetilde{U}(x,s)$,
i.e. 
\begin{equation}
\widetilde{U}(x,s)=\int_{0}^{L}s\, G(x,\xi,s)\frac{u(\xi,0)}{c^{2}}\, d\xi+\int_{0}^{L}G(x,\xi,s)\frac{\dot{u}(\xi,0)+p(\xi,s)}{c^{2}}\, d\xi+2\frac{h_{3}}{c}G(x,a,s)\, u(a,0).\label{eq:Ustilde}
\end{equation}
 As shown in Appendix, inhomogeneity in the boundary conditions introduces
two extra terms to the right hand side of Eq.\eqref{eq:Ustilde} analogous
to its last term, so 
\begin{equation}
U(x,s)=\widetilde{U}(x,s)+\frac{h_{1}}{c}G(x,0,s)\, u(0,0)+\frac{h_{2}}{c}G(x,L,s)\, u(L,0).\label{eq:Us}
\end{equation}
 Since $\Gamma(x,\xi,t):=\mathscr{L}^{-1}[G(x,\xi,s)]$ properties
of Laplace transform immediately imply 
\begin{align*}
\mathscr{L}^{-1}[s\, G(x,\xi,s)] & =\Gamma_{t}(x,\xi,t)\\
\mathscr{L}^{-1}[G(x,\xi,s)\, p(\xi,s)] & =\int_{0}^{t}\Gamma(x,\xi,t-\tau)\, p(\xi,\tau)\, d\tau.
\end{align*}
Technically speaking, $\mathscr{L}^{-1}[s\, G(x,\xi,s)]=\Gamma_{t}(x,\xi,t)+\Gamma(x,\xi,0)\,\delta(t)$,
but we have ignored the delta function since it does not contribute for $t>0$.
Therefore, solution to the initial-boundary problem Eqs.\eqref{eq:eom},\eqref{eq:eombc}
can be written in the form 
\begin{align}
u(x,t) & =\frac{1}{c}\Bigl[h_{1}u(0,0)\,\Gamma(x,0,t)+h_{2}u(L,0)\,\Gamma(x,L,t)+2h_{3}u(a,0)\,\Gamma(x,a,t)\Bigr]\label{eq:uxt}\\
 & \quad+\frac{1}{c^{2}}\int_{0}^{L}\Bigl[\Gamma_{t}(x,\xi,t)\, u(\xi,0)+\Gamma(x,\xi,t)\,\dot{u}(\xi,0)\Bigr]d\xi+\frac{1}{c^{2}}\int_{0}^{t}\int_{0}^{L}\Gamma(x,\xi,t-\tau)\, p(\xi,\tau)\, d\xi\, d\tau\notag
\end{align}
 If $h_{i}\neq\pm1$, $h_{1}+h_{2}+2h_{3}\neq0$ and all roots of
algebraic equation Eq.\eqref{eq:AlgCharEq} are simple then all our
assumptions for the validity of Eq.\eqref{eq:Gamt} are satisfied
and we can substitute it into Eq.\eqref{eq:uxt}. Taking into account
that 
\begin{align*}
\Gamma(x,0,t) & =\ \frac{c}{h_{1}+h_{2}+2h_{3}}\ +\ \sum_{k=1}^{p}\sum_{n=-\infty}^{\infty}\ \frac{1}{A_{n}^{(k)}}\,\varphi_{a}(x,s_{n}^{(k)})\, e^{s_{n}^{(k)}t}\\
\Gamma_{t}(x,\xi,t) & =\ \sum_{k=1}^{p}\sum_{n=-\infty}^{\infty}\ \frac{s_{n}^{(k)}}{A_{n}^{(k)}}\,\varphi_{a}(x,s_{n}^{(k)})\,\varphi_{a}(\xi,s_{n}^{(k)})\, e^{s_{n}^{(k)}t},
\end{align*}
 we have for the vibratory response: 
\begin{align}
u(x,t) & =\ \frac{1}{h_{1}+h_{2}+2h_{3}}\Bigg[h_{1}\, u(0,0)+h_{2}\, u(L,0)+2h_{3}\, u(a,0)+\frac{1}{c}\int_{0}^{L}\Bigl[\dot{u}(\xi,0)+\int_{0}^{t}p(\xi,\tau)\, d\tau\Bigr]d\xi\,\Bigg]\notag\label{eq:Fuxt}\\
 & \quad+\sum_{k=1}^{p}\sum_{n=-\infty}^{\infty}\ \frac{\varphi_{a}(x,s_{n}^{(k)})\, e^{s_{n}^{(k)}t}}{c^{2}A_{n}^{(k)}}\Bigg[\ \Bigl[c\, h_{1}\, u(0,0)+c\, h_{2}\, u(L,0)\,\varphi_{a}(L,s_{n}^{(k)})+2c\, h_{3}\, u(a,0)\,\varphi_{a}(a,s_{n}^{(k)})\Bigr]\\
 & \hspace{5.2cm}+\int_{0}^{L}\Bigl[s_{n}^{(k)}\, u(\xi,0)+\dot{u}(\xi,0)+\int_{0}^{t}p(\xi,\tau)\, e^{-s_{n}^{(k)}\tau}\, d\tau\Bigr]\varphi_{a}(\xi,s_{n}^{(k)})\, d\xi\ \Bigg].\notag
\end{align}
 Formula Eq.\eqref{eq:uxt} remains valid of course even if the eigenmode
expansion is inapplicable. For example, in those critical cases when
$\Gamma(x,\xi,t)$ is found in closed form, see Eqs.\eqref{eq:Gmh30h21},\eqref{eq:Gmh1h21},
we can use it to find the vibratory response as well. When substituting
into Eq.\eqref{eq:uxt} one should keep in mind that $H_{t}(t-\alpha)=\delta(t-\alpha)$,
where $\delta(t-\alpha)$ is the delta function of Dirac. Its convolution
with any function simply returns the function's value at $\alpha$.
Also, when terms like $H(\xi-a)H(ct-|x-\xi|)$ are integrated over
$\xi$ from $0$ to $L$ one has to consider cases $x>\xi$ and $x<\xi$
separately to remove the absolute value, e.g. 
\[
\int_{0}^{L}H(\xi-a)H(ct-|x-\xi|)\, w(\xi)\, d\xi=\int_{\max\{a,x-ct\}}^{x}\hspace{-0.7cm}w(\xi)\, d\xi+\int_{\max\{a,x\}}^{\min\{L,x+ct\}}w(\xi)\, d\xi
\]
Note that the first integral is assumed to vanish if its lower limit
exceeds its upper limit, e.g. $a>x$. Explicit expressions are cumbersome
and are given in the Appendix.

\section{Numerical issues}

\label{s9}

In this section we illustrate the behavior of the system for various
values of parameters $h_{1}$, $h_{2}$ and $h_{3}$. The critical cases are 
of particular interest as they demonstrate somewhat counter-intuitive behavior of the bar. 
Our analytical expressions are calculated using Maple and in some cases we compare results to
a MATLAB finite element implementation. We subject the bar to initial
displacement only in order to illustrate how each damper at the boundary
and along the bar modifies traveling waves. We use a Gaussian function
$0.1\,\exp(-(x-\mu)^{2}/(2\,\sigma^{2}))/(\sigma\,\sqrt{2\pi})$ as
an approximation of an impulse function where $\sigma=0.1$. We first
begin with some special values of the parameters $h_{i}$. For all
figures $c=1.5$ and $L=1.8$.

\subsection{Response of the system for $h_{2}=1$, $h_{3}=0$}

\begin{figure}[H]
 \centering{}\includegraphics[scale=0.4]{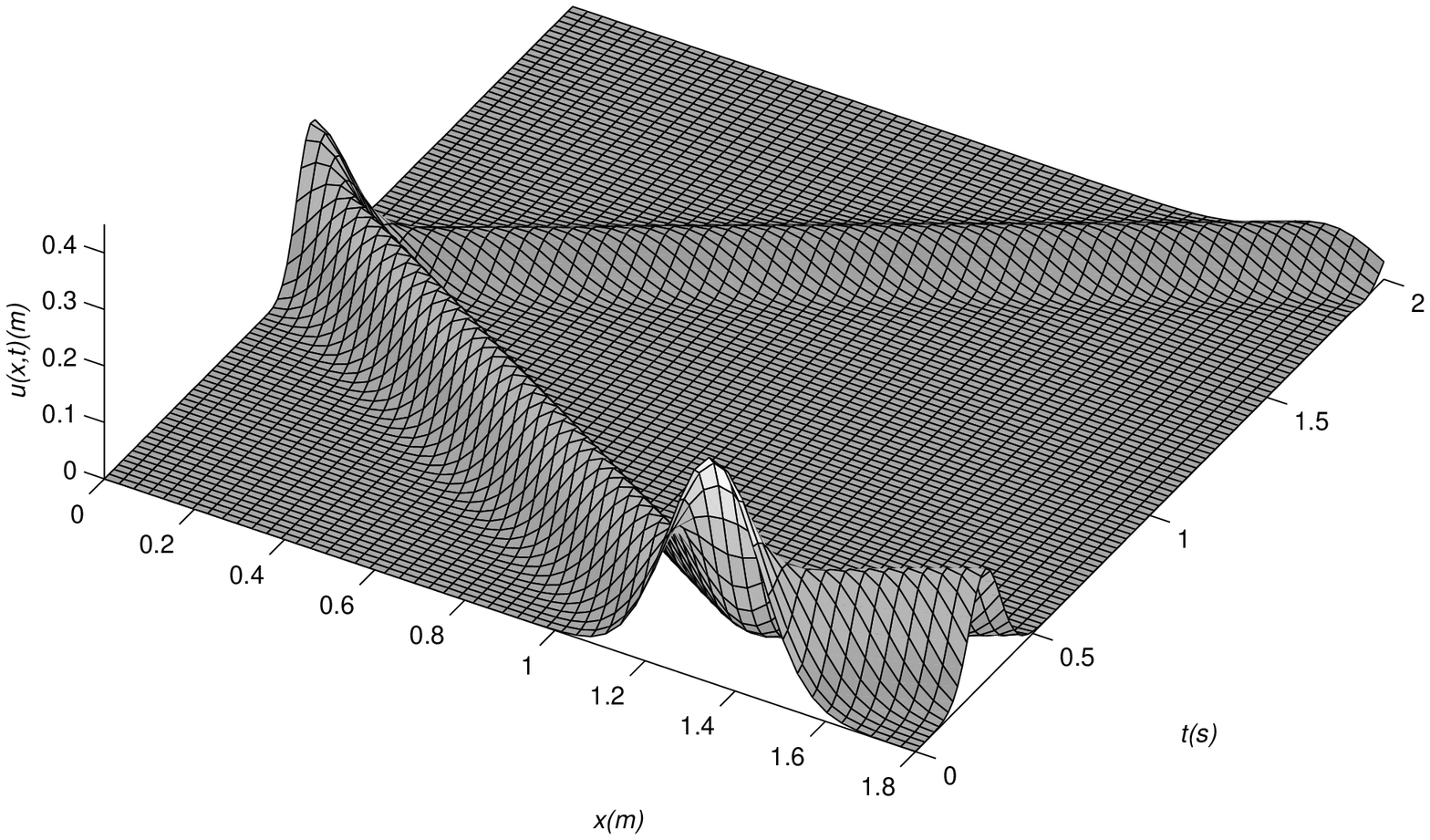}\includegraphics[scale=0.4]{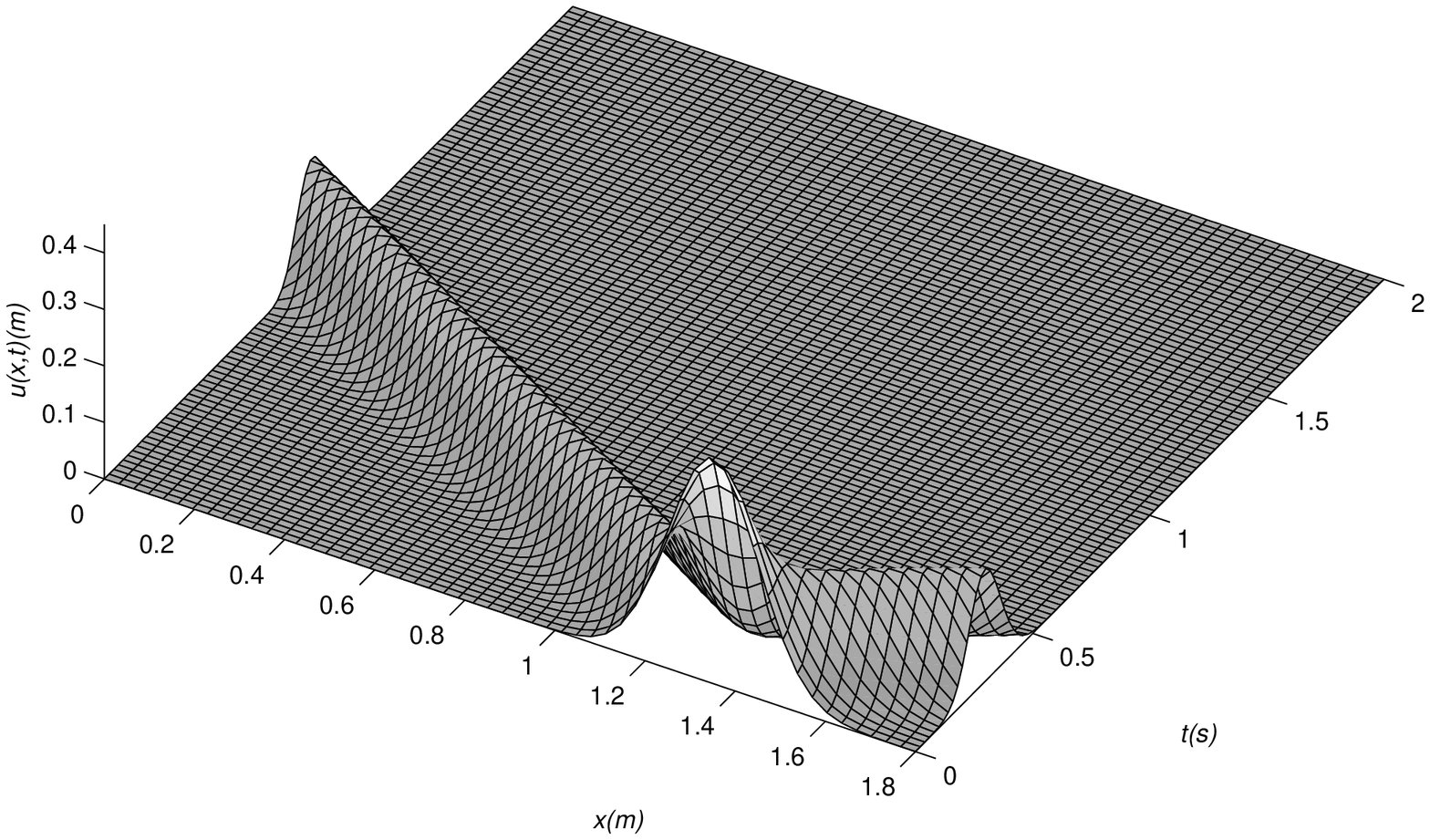}
\caption{\label{cap:supStab} Vibratory response for $h_{3}=0$, $h_{1}=0.5$,
$h_{2}=1$ (left), and $h_{1}=h_{2}=1$ (right).}
\end{figure}

Figure \ref{cap:supStab} shows the response of the system when $h_{1}=0.5$, 
$h_{2}=1$ and $h_{3}=0$. The right boundary is transparent and the left-half
of the traveling wave disappears on the right while the right-half
get reflected from the left boundary and then it too disappears on
the right. This behavior can be termed super-stable because the bar
comes to rest in finite time. Such a phenomenon can only be appropriately
observed using explicit solutions and it is the consequence of the
continuum nature of the system. Once we discretize the system this
behavior disappears and the bar could only come to rest at infinite
time. The reason for this is an exponential nature of the solution
for a discrete system. 

\subsection{Response of the system for $h_{1}=h_{2}=1$, $h_{3}\neq0$}

\begin{figure}[H]
\begin{centering}
\includegraphics[scale=0.6]{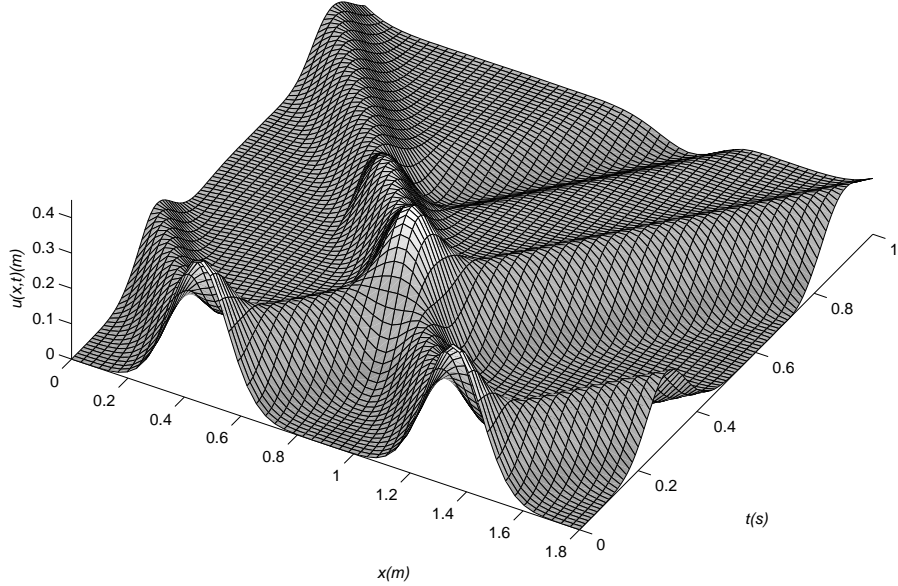} 
\par\end{centering}

\caption{Vibratory response for $h_{1}=h_{2}=1$ and $h_{3}=0.5$}

\label{cap:MidStab} 
\end{figure}

Figure \ref{cap:MidStab} depicts the response of the system for $h_{1}=h_{2}=1$,
$h_{3}=.5$. The internal damper is placed at distance $a=0.5$ from
the left end of the bar. The initial pulse consists of two Gaussian
functions centered at $0.25\, L$ and $0.75\, L$ . We see that a
wave gets partially reflected by the internal damper. This can be
best observed in the Figure 2 for the left-half of the right impulse.
For greater values of the $h_{3}$parameter the internal damper acts
almost as if it is the fixed point of the bar and waves get reflected
to a greater extent while only a small portion of the wave is passes
through. 

\subsection{Response of the system for $h_{2}=0.99$, $h_{3}\neq0$}

\begin{figure}[H]
\begin{centering}
 \includegraphics[scale=0.5]{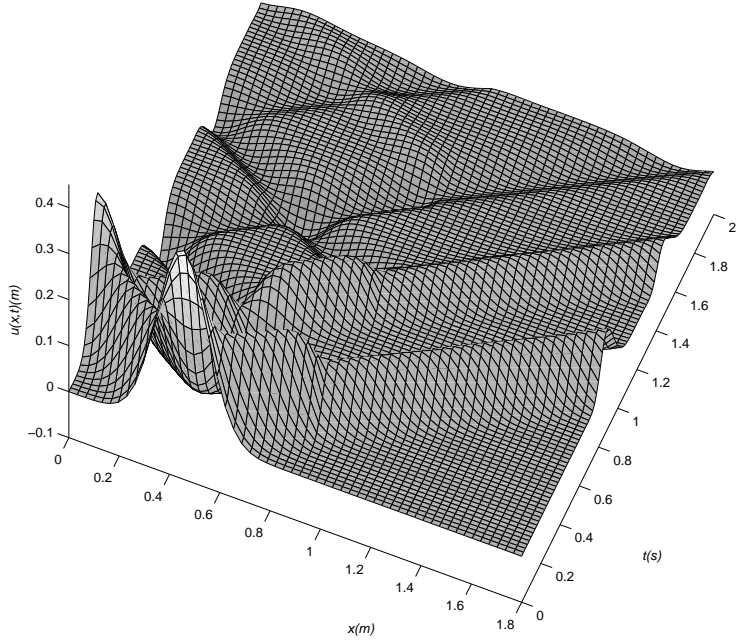}
\par\end{centering}
\caption{Vibratory response for $h_{1}=0.3$, $h_{2}=0.99$ and 
$h_{3}=0.7 .$}

\label{cap:NearStab} 
\end{figure}

Figure \ref{cap:NearStab} depicts the response of the system for
$h_{1}=0.3$, $h_{2}=0.99$ and $h_{3}=0.7$ using 40 eigenfunctions and a Gaussian
impulse at $\mu=0.25\, L$ . The internal damper is placed at distance
$a=0.5\, L$ from the left end of the bar. In this case the right
boundary is almost transparent which precludes formation of standing
waves on the right of the internal damper. The left boundary, however,
is reflective and the standing waves are clearly visible on Figure \ref{cap:NearStab}. This solution was validated with
a finite element implementation (FEM) calculated with 160 elements and the error was found to be within 0.001.
As the number of elements in FEM was increased this difference became
smaller indicating the efficiency of the analytical expression.

\subsection{Response of the system for $h_{1}h_{2}=1$ and $h_{3}=-\frac{h_{1}+h_{2}}{2}$}

\begin{figure}[H]
 \includegraphics[scale=0.4]{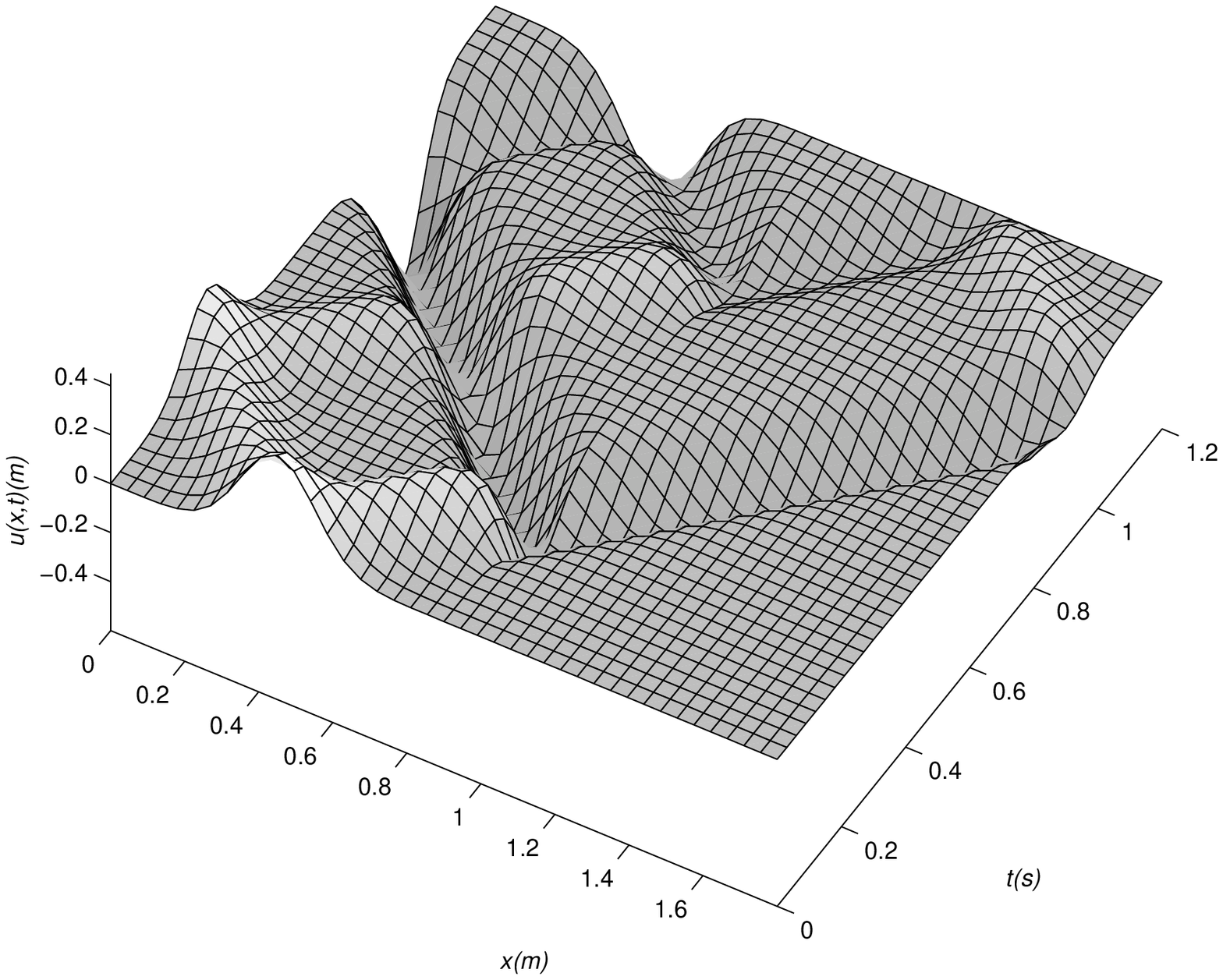}\includegraphics[scale=0.35]{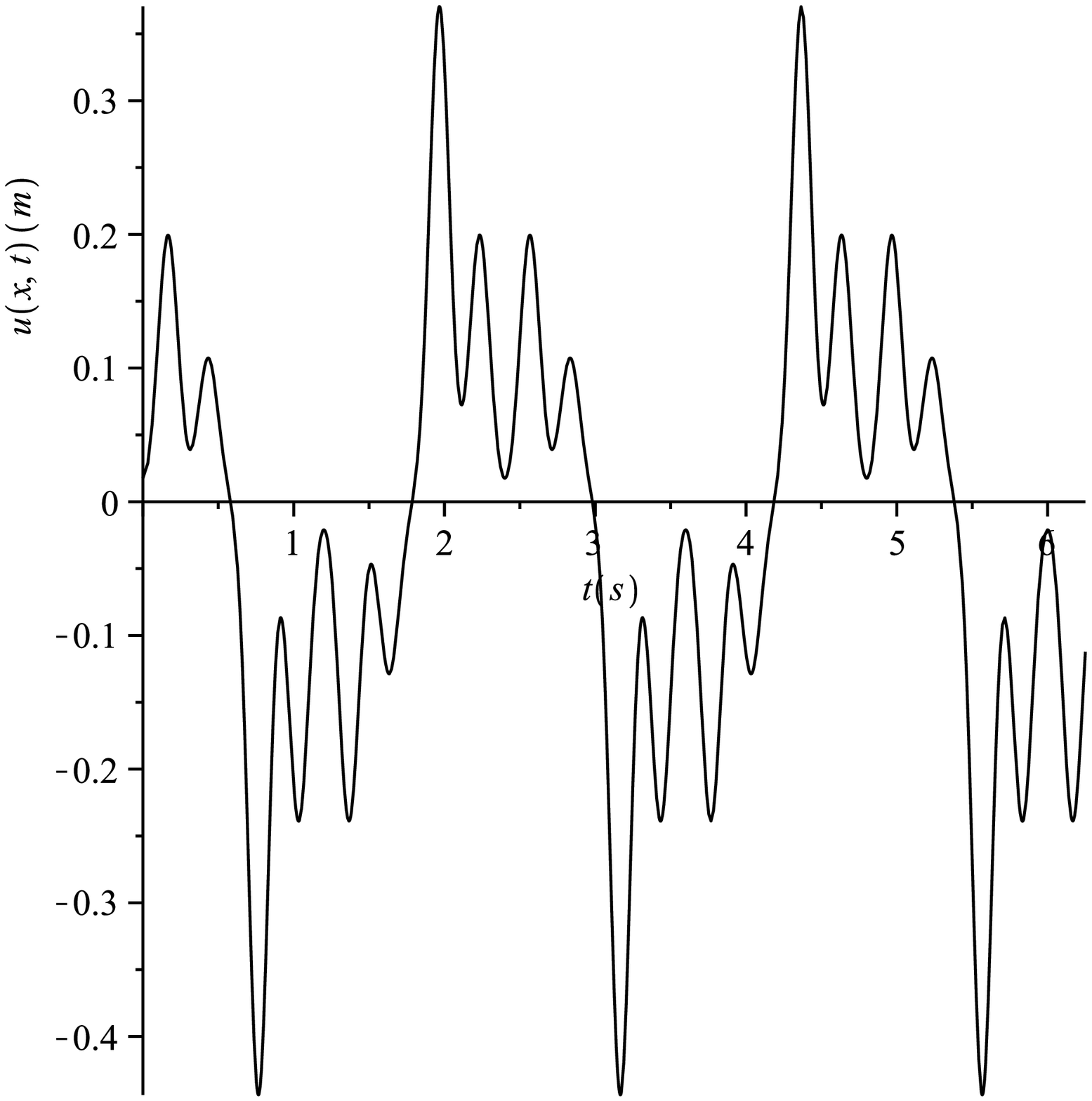}
\caption{$u(x,t)$ for $h_{1}=3/10$, $h_{2}=10/3$, $h_{3}=-109/60$ and $a=0.5\, L$(left)
and $u(\frac{1}{9}L,t)$ (right).}

\label{cap:unDamp} 
\end{figure}

Figure \ref{cap:unDamp} shows the response of the system for $h_{1}=3/10$,
$h_{2}=10/3$, $h_{3}=-109/60$ and $a=L/2$ with a Gaussian impulse
at $\mu=0.25\, L$. This is the case 3) from section \ref{sec:Damper-in-the}.
In this case the system has a double pole at zero and all the eigenvalues
are imaginary, i.e. there is no damping present in the bar. The amount
of energy lost at the left and right boundaries is returned into the
bar by the damper in the middle. Therefore, the displacement of every
point on the bar undergoes periodic motion as depicted in Figure \ref{cap:unDamp}.
Furthermore, this case is significant since FEM does not yield the
correct result. Our simulations indicate that the internal damper
greatly decreases the accuracy of FEM eigenvalue computation. As a
result, when all the poles are on the imaginary axis errors in the
real part significantly distort the response. Similar situation occurs
in case 4) of section \ref{sec:Damper-in-the}. However, in both cases
FEM response is even more distorted by spurious eigenvalues which we
discuss next. 

\subsection{Spurious eigenvalues in FEM}

\begin{figure}[H]
\begin{centering}
\includegraphics[scale=0.5]{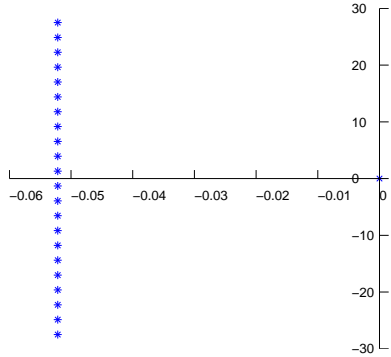} 
\par\end{centering}

\caption{Eigenvalues in the complex plane for $h_{1}=0.7$, $h_{2}=-1.5$.}

\label{cap:SpurEig} 
\end{figure}

We have encountered a phenomenon for which we do not have an explanation
at this time. We have observed that for some values of parameters
$h_{i}$ the stable continuous system becomes unstable when discretized
by the FEM method. One set of parameters that produces such a behavior
is $h_{1}=0.7$, $h_{2}=-1.5$ and $h_{3}=0$ for which the distribution
of its eigenvalues is depicted in Figure \ref{cap:SpurEig}. It is
clear that the continuous system is stable since there are no eigenvalues
with positive real parts. 

If, however, one discretizes the system an eigenvalue with a positive
real part will arise. It can be observed that a spurious eigenvalue
lies on the positive part of the real axis in the complex plane which
forces the discrete system to become unstable. The situation is not
improved by increasing the number of finite elements. On the contrary,
the real part of the spurious eigenvalue will increase to infinity
making the system even more unstable. This is counter-intuitive since
we know that the continuous system is equivalent to the discrete one
as the number of elements tends to infinity. Note that this phenomenon
has nothing to do with the system being unconstrained. Therefore, it
can be concluded that at least some stability regions for some parameters
of the continuous system become so distorted that after discretization
we observe unstable behavior. This phenomenon may have its roots in
the non-self-adjointness of the continuous system and at this time
it is unclear to us how a discretization of such a system changes
its behavioral pattern.

\section{Conclusions}

\label{s10}

We studied longitudinal vibrations of a bar with viscous ends and
internal damper. The corresponding eigenvalue problem contains the
spectral variable in the boundary conditions and has complex-valued,
non-orthogonal eigenmodes. Behavior of the system is controlled by
four dimensionless parameters, the three damping coefficients $h_{i}$,
and the relative position of the internal damper $a/L$. Our main
observations are summarized below.

Despite the unconventional nature of the eigenvalue problem the eigenmodes
can be found explicitly if the eigenvalues are known. When there is no internal damper or it is located in the middle of
the bar the eigenvalues can be found analytically as well. Otherwise,
for rational $a/L$ their determination reduces to solving an algebraic
equation. Distribution of eigenvalues is hypersensitive to the value
of $a/L$: its complexity grows swiftly with the denominator of $a/L$,
and it becomes pseudo-random for irrational $a/L$. 

When the values of $h_{i}$ are not restricted to be non-negative,
i.e. the dampers are allowed to be boosters, effective zero damping
may occur. Combinations of parameters that lead to zero damping are
non-trivial, but can be found analytically at least when there is
no internal damper or it is located in the middle of the bar. 

Generically, the eigenmodes are sufficient to expand the Green's function.
An explicit series solution can then be found for the vibratory response,
it is a superposition of exponentially damped (or boosted) standing
waves. These waves however, are complex-valued. Non-generic behavior occurs in two different situations. The first
one is that the eigenvalues may have higher multiplicity and the associated
modes are required for complete expansion. This only occurs when Eq.\eqref{eq:AlgCharEq}
has multiple roots or $h_{1}+h_{2}+2h_{3}=0$. In practice, one can
sidestep the issue by slightly perturbing the damping coefficients
to resolve the multiplicity. 

The second situation is critical behavior, when the eigenvalues disappear
partly or fully. This only happens for $h_{i}=\pm1$, but perturbing
the coefficients will lead to a qualitatively different picture at
large times. When $h_{i}=1$ the corresponding damper turns into a
perfect absorber draining all the energy from the bar in finite time.
With no reflection the standing waves either can not form at all (super-stability),
or can only form on a part of the bar accounting for scarcity of the
eigenmodes. When $h_{i}=-1$ one of the dampers turns into an infinite
amplifier and the solution can only exist before the traveling waves
reach that damper (super-instability). Mathematically, the eigenmode
expansion is impossible because the Laplace transformed Green's function
is unbounded at infinity and is not equal to the sum of partial fractions
over its poles. When no eigenmodes exist the solution can be found in closed from,
but partial cases like $h_{1}\neq1,h_{2}=1,h_{3}\neq0$ require further
work. Neither eigenmode expansion, nor closed form solution are possible
for such cases. This can only happen when the internal damper is present. 

FEM provides good approximation of the vibratory response for small
times even in critical cases. However, it is unreliable for determination
of eigenvalues even when the number of elements is large. Moreover,
it universally produces eigenvalues with large real parts that do
not converge to any actual eigenvalues as the number of elements is
increased. When these spurious eigenvalues have positive real parts
the FEM vibratory response is also unreliable for large times.

\section*{Appendix A: Coefficients of partial fractions}

\setcounter{equation}{0} \global\long\global\long\def\theequation{A.\arabic{equation}}

In this appendix we derive an explicit formula for the coefficients
of the partial function expansion of the Green's function. For simplicity,
within this Appendix we write $s_{n}$, $A_{n}$ instead of $s_{n}^{(k)}$,
$A_{n}^{(k)}$.

To this end consider the integral 
\begin{equation}
I:=\int_{0}^{L}G(x,\xi,s)\left(\frac{s^{2}}{c^{2}}\varphi_{a}(\xi,s_{n})-2h_{3}\frac{s}{c}\delta(\xi-a)\varphi_{a}(\xi,s_{n})-\varphi_{a}''(\xi,s_{n})\right)d\xi.\label{eq:I_starting_expr}
\end{equation}
 Taking into account that $G(\xi,x,s)$ and $\varphi_{a}(\xi,s_{n})$
both satisfy the boundary conditions the integral $I$ becomes after
integration by parts, 
\begin{equation}
I=\varphi_{a}(x,s_{n})+\frac{s-s_{n}}{c}\left[h_{1}G(x,0,s)\varphi_{a}(0,s_{n})+h_{2}G(x,L,s)\varphi_{a}(L,s_{n})\right].\label{eq:intbp}
\end{equation}
 Now we wish to take the limit $s\rightarrow s_{n}$. Contrary to
intuition, the second term does not tend to 0 when $s\rightarrow s_{n}$
because $G$ has a pole at $s_{n}$. Indeed, we have from Eq.\eqref{eq:G_expansion}
that $G(x,\xi,s)=\frac{\varphi_{a}(x,s_{n})\varphi_{a}(\xi,s_{n})}{A_{n}(s-s_{n})}+\text{Reg}(x,\xi,s)$
where $\text{Reg}(x,\xi,s)$ is regular at $s=s_{n}$. Substituting
this expression we find that 
\begin{eqnarray}
I & = & \varphi_{a}(x,s_{n})+\frac{1}{c}\left[h_{1}\frac{\varphi_{a}(x,s_{n})\varphi_{a}(0,s_{n})^{2}}{A_{n}}+h_{2}\frac{\varphi_{a}(x,s_{n})\varphi_{a}(L,s_{n})^{2}}{A_{n}}\right]+O(s-s_{n})\nonumber \\
 &  & \xrightarrow[s\to s_n]{}\varphi_{a}(x,s_{n})\left[1+\frac{1}{A_{n}c}\left(h_{1}\varphi_{a}(0,s_{n})^{2}+h_{2}\varphi_{a}(L,s_{n})^{2}\right)\right]\label{eq:I_1st_way}
\end{eqnarray}
On the other hand, we can compute $I$ in a different way. Since
$\varphi_{a}''(\xi,s_{n})=2h_{3}\frac{s_{n}}{c}\delta(\xi-a)\varphi_{a}(\xi,s_{n})+\frac{s_{n}^{2}}{c^{2}}\varphi_{a}(\xi,s_{n})$
from the equation we have 
\begin{equation}
2h_{3}\frac{s}{c}\delta(\xi-a)\varphi_{a}(\xi,s_{n})+\frac{s^{2}}{c^{2}}\varphi_{a}(\xi,s_{n})-\varphi_{a}''(\xi,s_{n})=2h_{3}\frac{s-s_{n}}{c}\delta(\xi-a)\varphi_{a}(\xi,s_{n})+\frac{s^{2}-s_{n}^{2}}{c^{2}}\varphi_{a}(\xi,s_{n}).\label{eq:main_eigen_s}
\end{equation}
Substituting Eq.\eqref{eq:main_eigen_s} into Eq.\eqref{eq:I_starting_expr}
yields 
\begin{eqnarray}
I & = & \int_{0}^{L}\left(\frac{\varphi_{a}(x,s_{n})\varphi_{a}(\xi,s_{n})}{A_{n}(s-s_{n})}+\text{Reg}(x,\xi,s_{n})\right)\left(2h_{3}\frac{s-s_{n}}{c}\delta(\xi-a)\varphi_{a}(\xi,s_{n})+\frac{s^{2}-s_{n}^{2}}{c^{2}}\varphi_{a}(\xi,s_{n})\right)d\xi\nonumber \\
 &  & \xrightarrow[s\to s_n]{}\varphi_{a}(x,s_{n})\left[\frac{2h_{3}}{A_{n}c}\int_{0}^{L}\delta(\xi-a)\varphi_{a}(L,s_{n})^{2}d\xi+\frac{2s_{n}}{A_{n}c^{2}}\int_{0}^{L}\varphi_{a}(\xi,s_{n})^{2}d\xi\right]\label{eq:I_2nd_way}
\end{eqnarray}
 Combining Eqs.\eqref{eq:I_1st_way} and \eqref{eq:I_2nd_way} finally
leads to 
\begin{equation}
A_{n}=\frac{2s_{n}}{c^{2}}\int_{0}^{L}\varphi_{a}(\xi,s_{n})^{2}d\xi+\frac{h_{1}}{c}\varphi_{a}(0,s_{n})^{2}+\frac{h_{2}}{c}\varphi_{a}(L,s_{n})^{2}+\frac{2h_{3}}{c}\varphi_{a}(a,s_{n})^{2}.
\end{equation}
 This formula is derived in \cite{OrShk} with missing boundary terms
because the authors overlook that the second term in Eq.\eqref{eq:intbp}
does not vanish as $s\to s_{n}$. After some algebra and simplifications
the formula is given by

\begin{multline}
A_{n}=-\frac{s_{n}}{c^{2}}\Bigg\{4\left(L-a\right)\varphi(a,s_{n})^{2}h_{3}+\Bigg[\left(2\,\left(L-\frac{1}{2}\, a\right)\left(h_{1}-1\right){{\rm e}^{-{\frac{s_{n}a}{c}}}}+2\,\left(h_{1}+1\right)\left(L-\frac{1}{2}\, a\right){{\rm e}^{{\frac{s_{n}a}{c}}}}\right)\varphi(a,s_{n})\\
+\frac{1}{2}\, a\left(h_{1}-1\right)^{2}{{\rm e}^{-2\,{\frac{s_{n}a}{c}}}}-\frac{1}{2}\, a\left(h_{1}+1\right)^{2}{{\rm e}^{2\,{\frac{s_{n}a}{c}}}}{\it \Bigg]}h_{3}+L(h_{1}^{2}-1)\Bigg\}.
\end{multline}

\section*{Appendix B: Inhomogeneous boundary conditions}

\setcounter{equation}{0} \global\long\global\long\def\theequation{B.\arabic{equation}}

Consider the boundary value problem 
\begin{equation}
\left(2h_{3}\frac{s}{c}\,\delta(x-a)+\frac{s^{2}}{c^{2}}\right)U(x,s)-\frac{d^{2}U(x,s)}{dx^{2}}=w(x,s)\label{eq:LaplTranMain}
\end{equation}
 
\begin{equation}
\frac{dU(0,s)}{dx}-\frac{h_{1}}{c}sU(0,s)=\gamma_{1}\label{eq:LaplTranMainBC1}
\end{equation}
 
\begin{equation}
\frac{dU(L,s)}{dx}+\frac{h_{2}}{c}sU(L,s)=\gamma_{2}\label{eq:LaplTranMainBC2}
\end{equation}
 Let $\widetilde{U}(x,s)$ be the solution with $\gamma_{1}=\gamma_{2}=0$.
Then $\widetilde{U}(x,s)=\int_{0}^{L}G(x,\xi,s)w(\xi,s)d\xi$, where
$G(x,\xi,s)$ is the Green's function of the homogeneous system. We
wish to find a solution for general $\gamma_{1},\,\gamma_{2}$.

The Green's function satisfies $L\left[G(x,\xi,s)\right]=\delta(x-\xi)$
with $L=\left(2h_{3}\frac{s}{c}\delta(x-a)+\frac{s^{2}}{c^{2}}\right)-\frac{d^{2}}{dx^{2}}$
and the homogeneous boundary conditions. Consider the Green's function
of the adjoint problem. Let $L^{*}=\left(2h_{3}\frac{\overline{s}}{c}\delta(x-a)+\frac{\overline{s}^{2}}{c^{2}}\right)-\frac{d^{2}}{dx^{2}}$
be the adjoint operator to $L$ \cite[chap 20.3]{Hassani}. The adjoint Green
function $g$ satisfies $L^{*}\left[g(x,\xi,s)\right]=\delta(x-\xi)$
and the adjoint boundary conditions.

We begin with the integration by parts to obtain 
\begin{equation}
\int_{0}^{L}\bar{g}(x,\xi,s)L\left[U(x,s)\right]dx=\left[\bar{g}(x,\xi,s)U_{x}(x,s)-\bar{g}_{x}(x,\xi,s)U(x,s)\right]_{0}^{L}+\int_{0}^{L}\overline{L^{*}g(x,\xi,s)}U(x,s)dx\label{eq:IntByPrts}
\end{equation}
 In view of the boundary conditions the first term on the right of
Eq.\eqref{eq:IntByPrts}, the so called surface term, simplifies to
$\bar{g}(L,\xi,s)\gamma_{2}-\bar{g}(0,\xi,s)\gamma_{1}$, while the
second term becomes $U(\xi,s)$ because of $\delta(x-\xi)$. This
yields 
\begin{equation}
U(\xi,s)=\int_{0}^{L}\bar{g}(x,\xi,s)L\left[U(x,s)\right]dx-\bar{g}(L,\xi,s)\gamma_{2}+\bar{g}(0,\xi,s)\gamma_{1}\label{eq:IntByPrtsSimpl}
\end{equation}
 It is known \cite[chap 20.3]{Hassani} that $\bar{g}(x,\xi,s)=G(\xi,x,s)$.
Substituting this into Eq.\eqref{eq:IntByPrtsSimpl} and interchanging
$x$ and $\xi$ we finally obtain 
\begin{equation}
U(x,s)=\widetilde{U}(x,s)+\gamma_{1}G(x,0,s)-\gamma_{2}G(x,L,s).\label{eq:uxs}
\end{equation}

\section*{Appendix C: Vibratory response for $h_{1}=h_{2}=1$}

\setcounter{equation}{0} \global\long\global\long\def\theequation{C.\arabic{equation}}

When $h_{1}=h_{2}=1$ it is possible to obtain a closed form solution
for the vibratory response by substituting Eq.\eqref{eq:Gmh1h21}
into Eq.\eqref{eq:uxt}. Here we give the final expressions for the
integrals from the second line of Eq.\eqref{eq:uxt}. To shorten the
formulas we set $f(\xi):=u(\xi,0)$ and $g(\xi):=\dot{u}(\xi,0)$:
\begin{multline*}
\frac{1}{c^{2}}\int_{0}^{L}\Gamma_{t}(x,\xi,t)\, f(\xi)\, d\xi=\Bigg[-\frac{h_{3}}{2(1+h_{3})}f(2a-x-tc)H\left(t-\frac{a-x}{c}\right)+\frac{1}{2}f(x-tc)\\
+\frac{1}{2}f(x+tc)H\left(\frac{a-x}{c}-t\right)+\frac{1}{2(1+h_{3})}f(x+tc)H\left(t-\frac{a-x}{c}\right)\Bigg]H(a-x)\\
+\Bigg[-\frac{h_{3}}{2(1+h_{3})}f(2a-x+tc)H\left(t-\frac{x-a}{c}\right)+\frac{1}{2}f(x+tc)+\frac{1}{2}f(x-tc)H\left(\frac{x-a}{c}-t\right)\\
+\frac{1}{2(1+h_{3})}f(x-tc)H\left(t-\frac{x-a}{c}\right)\Bigg]H(x-a)\,;
\end{multline*}
\begin{multline*}
\frac{1}{c^{2}}\int_{0}^{L}\Gamma(x,\xi,t)\, g(\xi)\, d\xi=\Bigg[-\frac{h_{3}}{2(1+h_{3})}\int_{0}^{t}g(2a-x-\tau c)H\left(\tau-\frac{a-x}{c}\right)d\tau+\frac{1}{2}\int_{0}^{t}g(x-\tau c)d\tau\\
+\frac{1}{2}\int_{0}^{t}g(x+\tau c)H\left(\frac{a-x}{c}-\tau\right)d\tau+\frac{1}{2(1+h_{3})}\int_{0}^{t}g(x+\tau c)H\left(\tau-\frac{a-x}{c}\right)d\tau\Bigg]H(a-x)\\
+\Bigg[-\frac{h_{3}}{2(1+h_{3})}\int_{0}^{t}g(2a-x+\tau c)H\left(\tau-\frac{x-a}{c}\right)d\tau+\frac{1}{2}\int_{0}^{t}g(x+\tau c)d\tau\\
+\frac{1}{2}\int_{0}^{t}g(x-\tau c)H\left(\frac{x-a}{c}-\tau\right)d\tau+\frac{1}{2(1+h_{3})}\int_{0}^{t}g(x-\tau c)H\left(\tau-\frac{x-a}{c}\right)d\tau\Bigg]H(x-a)\,;
\end{multline*}
\begin{multline*}
\frac{1}{c^{2}}\int_{0}^{t}\int_{0}^{L}\Gamma(x,\xi,t-\tau)\, p(\xi,\tau)\, d\xi\, d\tau=\frac{1}{2c(1+h_{3})}\Bigg[-h_{3}\int_{0}^{t-\frac{2a-x-\xi}{c}}\int_{0}^{a}p(\xi,\tau)d\xi d\tau\\
+(1+h_{3})\int_{0}^{t-\frac{x-\xi}{c}}\int_{0}^{x}p(\xi,\tau)d\xi d\tau+(1+h_{3})\int_{0}^{t-\frac{\xi-x}{c}}\int_{x}^{a}p(\xi,\tau)d\xi d\tau+\int_{0}^{t-\frac{\xi-x}{c}}\int_{a}^{L}p(\xi,\tau)d\xi d\tau\Bigg]H(a-x)\\
+\frac{1}{2c(1+h_{3})}\Bigg[\int_{0}^{t-\frac{x-\xi}{c}}\int_{0}^{a}p(\xi,\tau)d\xi d\tau-h_{3}\int_{0}^{t-\frac{x+\xi-2a}{c}}\int_{x}^{L}p(\xi,\tau)d\xi d\tau\\
+(1+h_{3})\int_{0}^{t-\frac{x-\xi}{c}}\int_{a}^{x}p(\xi,\tau)d\xi d\tau+\int_{0}^{t-\frac{\xi-x}{c}}\int_{a}^{L}p(\xi,\tau)d\xi d\tau\Bigg]H(x-a)\,.
\end{multline*}

\bibliographystyle{unsrt}
\bibliography{vibref}

\end{document}